\RequirePackage{ifpdf}
\ifpdf % We are running pdfTeX in pdf mode
\documentclass[pdftex]{sigma}
\else
\documentclass{sigma}
\fi

\numberwithin{equation}{section}

\begin{document}

\allowdisplaybreaks

\renewcommand{\thefootnote}{$\star$}

\renewcommand{\PaperNumber}{071}

\FirstPageHeading

\ShortArticleName{Hopf Maps,  Lowest Landau Level, and Fuzzy Spheres}

\ArticleName{Hopf Maps, Lowest Landau Level, and Fuzzy Spheres\footnote{This paper is a
contribution to the Special Issue ``Noncommutative Spaces and Fields''. The
full collection is available at
\href{http://www.emis.de/journals/SIGMA/noncommutative.html}{http://www.emis.de/journals/SIGMA/noncommutative.html}}}

\Author{Kazuki HASEBE}

\AuthorNameForHeading{K. Hasebe}

\Address{Kagawa National College of Technology,  Mitoyo, Kagawa 769-1192, Japan}
\Email{\href{mailto:hasebe@dg.kagawa-nct.ac.jp}{hasebe@dg.kagawa-nct.ac.jp}}

\ArticleDates{Received May 05, 2010, in f\/inal form August 19, 2010;  Published online September 07, 2010}

\Abstract{This paper is a review of monopoles, lowest Landau level, fuzzy spheres, and their mutual relations.
The Hopf maps of division algebras provide
a prototype relation between monopoles and fuzzy spheres.
Generalization of complex numbers to Clif\/ford al\-geb\-ra is exactly analogous to
 generalization of  fuzzy two-spheres to higher dimensional fuzzy spheres.
Higher dimensional fuzzy spheres have an interesting hierarchical structure made of ``compounds'' of lower dimensional spheres.
We give a physical interpretation for such particular structure of fuzzy spheres by utilizing Landau models in generic even dimensions.
 With Grassmann algebra, we also introduce a graded version of the Hopf map, and discuss its relation to fuzzy supersphere in context of  supersymmetric Landau model.}


\Classification{17B70; 58B34; 81V70}

\tableofcontents

\renewcommand{\thefootnote}{\arabic{footnote}}
\setcounter{footnote}{0}

\section{Introduction}

Fuzzy two-sphere introduced by Madore  \cite{madore1992} was a typical and  important fuzzy manifold %with non-trivial curvature
where particular notions of fuzzy geometry have been cultivated.
In subsequent explorations of fuzzy geometry, higher dimensional and supersymmetric generalizations of the fuzzy spheres were launched by Grosse et al.~\cite{hep-th/9602115,hep-th/9507074,math-ph/9804013}.
Furthermore, as well recognized now,
 string theory provides a~natural set-up for non-commutative geometry and non-anti-commutative geometry \cite{hep-th/9908142,hep-th/0302078,hep-th/0306226}. In particular,
classical solutions of matrix models with Chern--Simons term are identif\/ied with fuzzy two-spheres  \cite{hep-th/9910053}, fuzzy four-spheres  \cite{hep-th/9712105} and fuzzy superspheres  \cite{hep-th/0311005}.
Fuzzy manifolds also naturally appear in the context of  intersections of D-branes \cite{hep-th/9911136,hep-th/0306250,hep-th/050813}.
In addition to applications to physics,
 the fuzzy spheres themselves have intriguing mathematical structures.
As Ho and Ramgoolam showed in~\cite{Ho2002} and subsequently in~\cite{Kimura2002} Kimura investigated,  the commutative limit of (even-dimensional) fuzzy sphere takes the particular form\footnote{The odd-dimensional fuzzy spheres can be given by  $S_F^{2k-1}\simeq SO(2k)/(U(1)\otimes U(k-1))$  \cite{hep-th/0207111,hep-th/0309212}. For instance, $S_{F}^3\approx S^2\times S^2$.}:
\[
S_{F}^{2k}\simeq SO(2k+1)/U(k).
\]
From this coset representation, one may f\/ind, though $S_F^{2k}$ is called  $2k$-dimensional fuzzy sphere, its genuine dimension is {\it not} $2k$ but $k(k+1)$.
Furthermore, the fuzzy spheres can be expressed as the lower dimensional fuzzy sphere-f\/ibration over the sphere
\[
S_F^{2k}\approx  S^{2k}\times S_{F}^{2k-2}.
\]
Thus, $S_F^{2k}$ has ``extra-dimensions'' coming from the lower dimensional fuzzy sphere $S_F^{2k-2}$.
One may wonder why fuzzy spheres %$\it{have~to}$
have such extra-dimensions.
A mathematical explanation may go as follows.  Fuzzif\/ication is performed by replacing Poisson bracket with commutator on a manifold. Then, to fuzzif\/icate a manifold, the manifold has to have a symplectic structure to be capable to def\/ine  Poisson bracket. However, unfortunately, higher dimensional spheres~$S^{2k}$ \mbox{$(k\ge 2)$} do not accommodate symplectic structure, and their fuzzif\/ication is not straightforward. A possible resolution is to adopt a minimally extended symplectic
 manifold with spherical symmetry.
The coset $SO(2k+1)/U(k)$  suf\/f\/ices for the requirement.

Since detail mathematical treatments of fuzzy geometry have already been found in the excellent reviews
 \cite{hep-th/9801182,hep-th/0203259,arXiv:hep-th/0407007,Azumathesis,hep-th/0511114,Abethesis}, in this paper, we mainly focus on a physical approach for understanding of fuzzy spheres,  brought by the
 developments of higher dimensional quantum Hall ef\/fect (see as a review~\cite{hep-th/0606161} and references therein). The lowest Landau level (LLL) provides a nice way for physical understanding of fuzzy geometry (see for instance~\cite{hep-th/0306251}).
As we shall see in the context,  generalization of the fuzzy two-spheres is closely related to the generalization of the complex numbers in 19th century.
Quaternions  discovered by W.~Hamilton were the f\/irst generalization of complex numbers~\cite{Hamilton1844}.
Soon after the discovery, the other division algebra known as octonions was also found (see for instance~\cite{Baez2002}).
Interestingly, the division algebras are closely related to topological maps from sphere to sphere in dif\/ferent dimensions, i.e.\ the Hopf maps  \cite{Hopf1931,Hopf1935}. While the division algebras consist of
 complex numbers $\mathbb{C}$, quaternions $\mathbb{H}$ and octonions   $\mathbb{O}$ (except for real numbers $\mathbb{R}$),
there is another generalization of complex numbers and quaternions; the celebrated Clif\/ford algebra invented by W.~Clif\/ford~\cite{Clifford1878}.
 The generalization of complex numbers to Clif\/ford algebra is
exactly analogous to generalization of fuzzy two-sphere to its higher dimensional cousins.
Particular geometry of fuzzy spheres directly ref\/lects features of Clif\/ford algebra.
With non-Abelian monopoles in generic even dimensional space, we explain the  particular geometry of fuzzy spheres in view of the lowest Landau level physics.
There is another important algebra invented by H.~Grassmann \cite{Grassmann1844}. Though less well known compared to the original three Hopf maps, there also exists a graded version of the (1st) Hopf map~\cite{LandiMarmo1987}.
  We also discuss its relation to fuzzy supersphere.

 The organization is as follows. In Section~\ref{hopfmapdivisionalgebra}, we
 introduce the division algebras and  the Hopf maps.
 In Section~\ref{1sthoppfandfuzzy2sphere}, with the explicit construction of the 1st Hopf map,  we analyze the Landau problem on a two-sphere and discuss its relations to fuzzy two-sphere $S_F^2$.
The graded version of the Hopf map and its relation to fuzzy superspheres are investigated in Section~\ref{sectsuperhopf}.
In Section~\ref{Sec2ndHopf}, we extend the former discussions to  the 2nd Hopf map and fuzzy four-sphere $S_F^4$.
In Section~\ref{sec3rdhopf}, we consider the 3rd Hopf map and the corresponding fuzzy manifolds.
In Section~\ref{sechbeyondhopf}, we generalize the observations to even higher dimensional fuzzy spheres based on Clif\/ford algebra.
Section~\ref{secsummary} is devoted to summary and discussions.
In Appendix~\ref{sect0thhopf}, for completeness, we introduce the 0th Hopf map and related ``Landau problem'' on a circle. In Appendix~\ref{sectioncomplexprojective}, the $SU(k+1)$ Landau model and
  fuzzy $\mathbb{C}P^k$ manifold are surveyed.

\section{Hopf maps and division algebras}\label{hopfmapdivisionalgebra}

As the division algebras consist of only three algebras, there exist  three corresponding Hopf maps, 1st, 2nd and 3rd. The three Hopf maps
\begin{center}
\begin{tabular}{ccccccc}
 & & $S^{3}$ &   $\overset{S^{1}}\longrightarrow $ & $S^{2}$ & & ~~~~~~~~~~(1st)\\
 &  $S^{7}$ & $\longrightarrow$ & $S^{4}$ &  & & ~~~~~~~~~~(2nd) \\
 $S^{15}$ & $\longrightarrow$ &   $S^{8}$ &&  & & ~~~~~~~~~~(3rd)
\end{tabular}
\end{center}
 are closely related
bundle structures of $U(1)$, $SU(2)$, and $SO(8)$ monopoles \cite{Dirac1931,Yang1978,Grossman1984}.
Interes\-ting\-ly, the Hopf maps exhibit a hierarchical structure.
Each of the Hopf maps can be understood as a map from a circle in 2D division algebra space to  corresponding projective space:
\begin{center}
\begin{tabular}{ccccccc}
 & & $S^{1}_{\mathbb{C}}$ &   $\overset{S^{1}}\longrightarrow $ & $\mathbb{C}P^{1}$ & & ~~~~~~~~~~(1st)\\
 &  $S^{1}_{\mathbb{H}}$ & $\longrightarrow$ & $\mathbb{H}P^{1}$ &  & & ~~~~~~~~~~(2nd) \\
 $S^{1}_{\mathbb{O}}$ & $\longrightarrow$ &   $\mathbb{O}P^{1}$ &&  & & ~~~~~~~~~~(3rd)
\end{tabular}
\end{center}
For instance, in the 1st Hopf map, the total space $S^1_{\mathbb{C}}$ represents a  circle in 2D complex space, i.e.~$S^3$, and the basespace $\mathbb{C}P^1$ denotes the complex projective space equivalent to $S^2$.   For the 2nd and 3rd  Hopf maps, same interpretations hold by replacing complex numbers $\mathbb{C}$ with quaternions $\mathbb{H}$, and octonions  $\mathbb{O}$, respectively.
 For later convenience, we summarize the basic properties of  quaternions  and octonions.
The quaternion basis elements,~$1$, $q_1$, $q_2$, $q_3$, are def\/ined so as to satisfy the algebra~\cite{Hamilton1844}
\begin{gather*}
 q_1^2=q_2^2=q_3^2=q_1q_2q_3=-1,\qquad
 q_i q_j=-q_jq_i \quad (i\neq j),
\end{gather*}
or equivalently,
\begin{gather*}
 \{q_i,q_j\}=-2\delta_{ij},\qquad
 [q_i,q_j]=2\epsilon_{ijk}q_k,
\end{gather*}
where $\epsilon_{ijk}$ is Levi-Civita antisymmetric tensor with $\epsilon_{123}=1$.
Thus, quaternion algebra is non-commutative.
As is well known, the quaternion algebra is satisf\/ied by the Pauli matrices with  the identif\/ication,
$q_i=-i\sigma_i$.  (We will revisit this point in Section~\ref{Sec2ndHopf}.)
An arbitrary quaternion is expanded by the quaternion basis elements:
\[
q=r_0 1+\sum_{i=1}^3 r_i q_i,
\]
where $r_0$, $r_i$ are real expansion coef\/f\/icients, and
the conjugation of $q$ is given by
\[
q^*=r_0 1-\sum_{i=1}^3 r_i q_i.
\]
The norm of quaternion, $||q||$, is given by
\[
||q||=\sqrt{q^*q}=\sqrt{qq^*}=\sqrt{{r_0}^2+\sum_{i=1}^3{r_i}^2}.
\]
It is noted that $q$ and $q^*$ are commutative.
The normalized quaternionic space corresponds to~$S^{3}$, and the total manifold of the 2nd Hopf map, $S^7$, is expressed as the quaternionic circle,~$S^1_{\mathbb{H}}$:
\[
q^* q+{q'}^* q'=r_0^2+\sum_{i=1}^3 r_i^2+{r'_0}^2+\sum_{i=1}^3{r'_i}^2=1.
\]
Similarly, the octonion basis elements,
$1,e_1,e_2, \dots,e_7$,  are def\/ined so as to satisfy the algebras
\begin{gather}
 \{e_I,e_J\}=-2\delta_{IJ},\qquad
 [e_I,e_J]=2f_{IJK}e_K,
\label{splitoctonionalgebra}
\end{gather}
where $I,J,K=1,2, \dots,7$. $\delta_{IJ}$ denotes Kronecker delta symbol and $f_{IJK}$ does the antisymmetric structure constants of octonions
(see Table~\ref{Octoniontable}).

\begin{table}[t]\centering
 \caption{The structure constants of the octonion algebra. For instance, $f_{145}=1$ can be read from $e_1 e_4 =e_5$. The octonion structure constants are given by  $f_{123}=f_{145}=f_{176}=f_{246}=f_{527}=f_{374}=f_{365}=1$, and other non-zero octonion structure constants are obtained by the cyclic permutation of the indices. }
 \label{Octoniontable}

 \vspace{1mm}

\begin{tabular}{|c||c|c|c|c|c|c|c|c|}
\hline       & 1  &  $e_1$     & $e_2$ & $e_3$  & $e_4$  & $e_5$ & $e_6$ & $e_7$ \\
\hline
\hline 1     & $1$  & $e_1$   & $e_2$  & $e_3$  & $e_4$  & $e_5$  & $e_6$  & $e_7$  \\
\hline $e_1$ & $e_1$ & $-1$    & $e_3$  & $-e_2$ & $e_5$  & $-e_4$  & $-e_7$ & $e_6$ \\
\hline $e_2$ & $e_2$ & $-e_3$  & $-1$   &  $e_1$ & $e_6$ & $e_7$  & $-e_4$  & $-e_5$  \\
\hline $e_3$ & $e_2$ & $e_2$   & $-e_1$ & $-1$   & $e_7$ & $-e_6$ & $e_5$  & $-e_4$   \\
\hline $e_4$ & $e_4$ & $-e_5$ & $-e_6$  & $-e_7$  & $-1$    & $e_1$  & $e_2$  & $e_3$ \\
\hline $e_5$ & $e_5$ & $-e_4$  & $-e_7$ & $e_6$  & $-e_1$ & $-1$      & $-e_3$  & $e_2$ \\
\hline $e_6$ & $e_6$ & $e_7$   & $e_4$ & $-e_5$ & $-e_2$ & $e_3$ & $-1$      & $-e_1$  \\
\hline $e_7$ & $e_7$ & $-e_6$  & $e_5$  & $e_4$ &$-e_3$  & $-e_2$  & $e_1$ & $-1$ \\
\hline
\end{tabular}
\end{table}

The octonions do not respect the associativity as well as the commutativity.
(The non-associativity can be read from Table~\ref{Octoniontable}, for instance, $(e_1e_2)e_4=e_7=-e_1(e_2e_4)$.)
Due to their non-associativity, octonions cannot be represented by matrices  unlike  quaternions. However, the conjugation and magnitude of octonion can be similarly def\/ined as those of quaternion, simply replacing the role of the imaginary quaternion basis elements $q_i$ with the imaginary octonion basis elements $e_I$.
An arbitrary octonion is given by
\[
o=r_0 1+\sum_{I=1}^7 r_I e_I,
\]
with real expansion coef\/f\/icients $r_0$, $r_I$, and  its conjugation is
\[
o^*=r_0 1-\sum_{I=1}^7 r_I e_I.
\]
The norm of octonion, $||o||$, is
\[
||o||=\sqrt{o^*o}=\sqrt{oo^*}=\sqrt{{r_0}^2+\sum_{I=1}^7{r_I}^2}.
\]
Like the case of quaternion, $o$ and $o^*$ are commutative.
The normalized octonion $o^*o=1$ represents $S^{7}$, and  the total manifold of the 3rd Hopf map,  $S^{15}$, is given by the octonionic circle~$S^1_\mathbb{O}$,
\[
o^* o+{o'}^* o'=r_0^2+\sum_{I=1}^7 r_I^2+{r'_0}^2+\sum_{I=1}^7{r'_I}^2=1.
\]
One may wonder there might exist even higher dimensional generalizations. Indeed,  following to the Cayley--Dickson construction \cite{Baez2002}, it is possible to  construct new species of numbers.
Next to the octonions,  sedenions consisting of 16 basis elements can be  constructed.
However, the sedenions do not even respect the alternativity, and hence   the multiplication law of norms does not hold: $||x||\, ||y||\neq  ||x \cdot y||$.
 Then, the usual concept of ``length'' does not even exist in sedenions, and a sphere cannot be def\/ined with sedenions and hence the corresponding Hopf maps either.
Consequently, there only exit  three division algebras and  corresponding three Hopf maps.

\section{1st Hopf map and fuzzy two-sphere}\label{1sthoppfandfuzzy2sphere}

In this section, we give a realization of the 1st Hopf map
\begin{gather*}
S^3 \overset{S^1}\longrightarrow S^2,
%\label{1stcopacthopf}
\end{gather*}
and discuss basic procedure of fuzzif\/ication of sphere in LLL.

\subsection[1st Hopf map and $U(1)$ monopole]{1st Hopf map and $\boldsymbol{U(1)}$ monopole}

The 1st Hopf map can explicitly be constructed as follows.
We f\/irst introduce a normalized complex two-spinor
\[
\phi=
\begin{pmatrix}
\phi_1\\
\phi_2
\end{pmatrix}
\]
satisfying $\phi^{\dagger}\phi=1$.
Thus  $\phi$, which we call the 1st Hopf spinor, represents the coordinate on the total space $S^3$, and plays a primary role in the fuzzif\/ication of sphere as we shall see below.
With the Pauli matrices
\begin{gather}
\sigma_1=
\begin{pmatrix}
0 & 1 \\
1 & 0
\end{pmatrix},
\qquad \sigma_2=
\begin{pmatrix}
0 & -i \\
i & 0
\end{pmatrix},\qquad
\sigma_3=
\begin{pmatrix}
1 & 0 \\
0 & -1
\end{pmatrix},
\label{paulimatrices}
\end{gather}
the f\/irst Hopf map is realized as
\begin{gather}
\phi\rightarrow x_i=\phi^{\dagger}\sigma_i\phi.
\label{1stHopfmapexpli}
\end{gather}
%%%%%%%%%%%%%%%%%%%%%%%%%%
It is straightforward to check that  $x_i$ satisfy the condition for $S^2$:
\begin{gather}
x_i x_i=(\phi^{\dagger}\phi)^2=1.
\label{xsquareclass}
\end{gather}
Thus, (\ref{1stHopfmapexpli}) demonstrates the 1st Hopf map.
The analytic form of the Hopf spinor except for the south pole is given by
\begin{gather}
\phi=\frac{1}{\sqrt{2(1+x_3)}}
\begin{pmatrix}
1+x_3 \\
x_1+ix_2
\end{pmatrix},
\label{1sthopfspinornorth}
\end{gather}
and the corresponding f\/ibre-connection is derived as
\begin{gather}
A=dx_i A_i=-i\phi^{\dagger}{d}\phi,
\label{oneform1stHopf}
\end{gather}
where
\[
A_i=-\frac{1}{2(1+x_3)}\epsilon_{ij3}x_j.
\]
The curvature is given by
\begin{gather}
F_{ij}=\partial_i A_j-\partial_j A_i=\frac{1}{2}\epsilon_{ijk}x_k,
\label{U1gaugefields}
\end{gather}
which corresponds to the f\/ield strength of Dirac monopole with minimum charge (see for instance~\cite{Nakaharabook}).
The analytic form of the Hopf spinor except for the north pole is given by
\begin{gather*}
\phi'=\frac{1}{\sqrt{2(1-x_3)}}
\begin{pmatrix}
x_1-ix_2 \\
1-x_3
\end{pmatrix}.
%\label{1sthopfspinorsouth}
\end{gather*}
The corresponding gauge f\/ield is
\[
A'=-i{\phi'}^{\dagger}d\phi'=dx_i A'_i,
\]
where
\[
A'_i=\frac{1}{2(1-x_3)}\epsilon_{ij3}x_j.
\]
The f\/ield strength $F'_{ij}=\partial_i A_j'-\partial_j A'_i$ is same as $F_{ij}$ (\ref{U1gaugefields}), which suggests the two expressions of the
 Hopf spinor % (\ref{1sthopfspinornorth}) and (\ref{1sthopfspinorsouth})
 are related by  gauge transformation. Indeed,
\[
\phi'=\phi \cdot g =g \cdot \phi,
\]
where $g$ is a $U(1)$ gauge group element
\[
g=e^{-i \chi}=\frac{1}{\sqrt{1-{x_3}^2}}(x_1-ix_2).
\]
Here, the gauge parameter $ \chi$ is given by $\tan ( \chi)=\frac{x_2}{x_1}$. The $U(1)$ phase factor is canceled in the map~(\ref{1stHopfmapexpli}), and there always exists such~$U(1)$ gauge degree of freedom in expression of the Hopf spinor.
With the formula
\[
-ig^*dg=\frac{1}{1-{x_3}^2}\epsilon_{ij3}x_idx_j,
\]
the above gauge f\/ields are represented as
\[
 A=\frac{i}{2}(1-x_3)g^*dg,\qquad
 A'=-\frac{i}{2}(1+x_3)g^*dg,
\]
and related by the $U(1)$ gauge transformation
\[
A'=A-ig^*dg.
\]
The non-trivial bundle structure of  $U(1)$ monopole on $S^2$ is guaranteed by the homotopy theorem
\[
\pi_1(U(1))\simeq \mathbb{Z},
\]
specif\/ied by the 1st Chern number
\[
c_1=\frac{1}{4\pi}\int_{S^2} F.
\]

\subsection[$SO(3)$ Landau model]{$\boldsymbol{SO(3)}$ Landau model}

Here, we consider a Landau model on a two-sphere in $U(1)$ monopole background.
In 3D space, the Landau Hamiltonian is given by
\begin{gather}
H=-\frac{1}{2M}D_i^2=-\frac{1}{2M}\frac{\partial^2}{\partial r^2}-\frac{1}{Mr}\frac{\partial}{\partial r}+\frac{1}{2Mr^2}\Lambda_i^2,
\label{3DLandauHamilt}
\end{gather}
where $D_i=\partial_i+iA_i$,   $r=\sqrt{x_ix_i}$, and $i=1,2, 3$ are summed over.   $\Lambda_i$ is the covariant angular momentum $\Lambda_i=-i\epsilon_{ijk}x_jD_k$.
The monopole gauge f\/ield has the form
\[
A_i=-\frac{I}{2r(r+x_3)}\epsilon_{ij3}x_j,
\]
and the corresponding f\/ield strength is
\[
F_i=\epsilon_{ijk}\partial_j A_k=\frac{I}{2r^3}x_i.
\]
The covariant angular momentum $\Lambda_i$ does not satisfy the $SU(2)$ algebra, but satisf\/ies
\[
[\Lambda_i,\Lambda_j]=i\epsilon_{ijk}\big(\Lambda_k-r^2F_k\big).
\]
The conserved $SU(2)$ angular momentum is constructed as
\[
L_i=\Lambda_i+r^2F_i,
%\label{conservedSU(2)angular}
\]
which satisf\/ies the genuine $SU(2)$ algebra
\[
[L_i,L_j]=i\epsilon_{ijk}L_k.
\]
On a two-sphere, the Hamiltonian (\ref{3DLandauHamilt}) is reduced to the $SO(3)$ Landau model~\cite{Haldane1983}
\begin{gather}
H=\frac{1}{2MR^2}\Lambda_i^2,
\label{on2DsphereLandau}
\end{gather}
where $R$ is the radius of two-sphere.
 The $SO(3)$  Landau Hamiltonian can be rewritten as
\begin{gather}
H=\frac{1}{2MR^2}\left(L_i^2-\frac{I^2}{4}\right),
\label{so3landauhamidef}
\end{gather}
 where we  used the orthogonality between the covariant angular momentum and the f\/ield\linebreak  strength; $\Lambda_i F_i =F_i \Lambda_i =0$.
Thus, the Hamiltonian is represented by the $SU(2)$ Casimir opera\-tor, and the eigenvalue problem is boiled down to the problem of obtaining the irreducible representation of~$SU(2)$. Such irreducible representations are given by monopole harmonics~\cite{monopoleharmonics1976}.
The $SU(2)$ Casimir operator takes the eigenvalues $L_i^2=j(j+1)$ with $j=\frac{I}{2}+n$ $(n=0,1,2, \dots)$. (The minimum of $j$ is not zero but a f\/inite value $I/2$ due to the existence of the f\/ield angular momentum of monopole.) Then, the eigenenergies are derived as
\begin{gather}
E_n=\frac{1}{2MR^2}\left(n^2+n(I+1)+\frac{I}{2}\right).
\label{energylevelsphere}
\end{gather}
In the thermodynamic limit, $R,I\rightarrow\infty$, with $B=I/2R^2$ f\/ixed,   (\ref{energylevelsphere}) reproduces the usual Landau levels on a plane:
\[
E_n\rightarrow \omega\left(n+\frac{1}{2}\right),
\]
where $\omega={B}/{M}$ is the cyclotron frequency.
 The degeneracy of the $n$th Landau level is given by
\begin{gather}
d(n)=2l+1=2n+I+1.
\label{degeneso3n}
\end{gather}
In particular, in the LLL $(n=0)$, the degeneracy is
\[
d_{\rm LLL}=I+1.
\]
The monopole harmonics in LLL is simply constructed by taking   symmetric products of the components of the 1st Hopf spinor
\begin{gather}
\phi_{\rm LLL}^{(m_1,m_2)}=\sqrt{\frac{I!}{m_1! m_2!}} \phi_1^{m_1}\phi_2^{m_2},
\label{LLLbasistwosphere}
\end{gather}
where $m_1+m_2=I$ ($m_1,m_2\ge 0$).
In the LLL, the kinetic term of the covariant angular momentum is quenched, and
the $SU(2)$ total angular momentum is reduced to the monopole f\/ield strength
\[
L_i \rightarrow r^2F_i= \frac{I}{2R}x_i.
\]
Then in LLL,  the coordinates $x_i$ are regarded as the operator
\[
X_i=\alpha L_i,
\]
which satisf\/ies the def\/inition algebra of fuzzy two-sphere{\samepage
\begin{gather}
[X_i,X_j]=i\alpha \epsilon_{ijk}X_k,
\label{fuzzytwospherealgera}
\end{gather}
with $\alpha=2R/I$.}

We reconsider the LLL physics  with Lagrange formalism. In Lagrange formalism,
importance of the Hopf map becomes more transparent. The present one-particle Lagrangian on  a two-sphere is given by
\[
L=\frac{m}{2} \dot{x}_i\dot{x}_i+\dot{x}_i A_i,
\]
with the constraint
\begin{gather}
x_i x_i=R^2.
\label{xsquareR}
\end{gather}
In the LLL, the Lagrangian is represented only by the interaction term
\[
L_{\rm LLL}=\dot{x}_i A_i.
\]
From (\ref{oneform1stHopf}), the LLL Lagrangian can be simply rewritten as
\begin{gather}
L_{\rm LLL}=-iI\phi^{\dagger}\frac{d}{dt}\phi,
\label{LLLlagrangiansu2}
\end{gather}
and the constraint (\ref{xsquareR}) is
\begin{gather}
\phi^{\dagger}\phi=1.
\label{constraintphinorma}
\end{gather}
It is noted that, in the LLL, the kinetic term drops and only the f\/irst order time derivative term survives. The Lagrangian and the constraint can be represented in terms of the Hopf spinor.
Usually, in the LLL, the quantization is preformed by regarding the Hopf spinor as  fundamental quantity, and the canonical quantization condition is imposed not on the original coordinate on two-sphere, but on the Hopf spinor.  We follow such quantization procedure to fuzzif\/icate   two-sphere. From (\ref{LLLlagrangiansu2}),
the conjugate momentum is derived as $\pi=-iI\phi^*$, and the canonical quantization is given by
\begin{gather}
[\phi_{\alpha},\phi_{\beta}^*]=-\frac{1}{I}\delta_{\alpha\beta}.
\label{canonicalcondphi}
\end{gather}
This quantization may remind the quantization procedure of the spinor f\/ield theory, but  readers should not be confused:  The present quantization is for one-particle  mechanics, and the spinor is quantized as ``boson'' (\ref{canonicalcondphi}).
Then, the complex conjugation is regarded as the derivative
\begin{gather}
\phi^*=\frac{1}{I}\frac{\partial}{\partial \phi},
\label{complexderivative}
\end{gather}
and the constraint (\ref{constraintphinorma}) is considered as a condition on the LLL basis
\begin{gather*}
\phi^t \frac{\partial}{\partial \phi}  \phi_{\rm LLL}=I \phi_{\rm LLL}.
%\label{constraintonstateLLL}
\end{gather*}
The previously derived LLL basis (\ref{LLLbasistwosphere}) indeed satisf\/ies the condition.

By inserting the expression (\ref{complexderivative}) to the Hopf map, we f\/ind that $x_i$ are regarded as  coordinates on  fuzzy two-sphere
\[
X_i =R\phi^{\dagger}\sigma_i \phi=\frac{\alpha}{2}\phi^t\sigma_i\frac{\partial}{\partial\phi},
\]
which satisf\/ies the algebra (\ref{fuzzytwospherealgera}).
Thus, also in the Lagrange formalism, we arrive at the fuzzy two-sphere algebra in   LLL. The crucial role of the Hopf spinor is transparent
in the Lagrange formalism:
The Hopf spinor is f\/irst fuzzif\/icated~(\ref{canonicalcondphi}), and subsequently the coordinates on
two-sphere are fuzzif\/icated. This is the basic fuzzif\/ication mechanics
of  sphere in the context of the Hopf map.

\subsection{Fuzzy two-sphere}

The fuzzy two-sphere is a fuzzy manifold  whose coordinates satisfy the $SU(2)$ algebraic relation~\cite{madore1992}
\[
[\hat{X}_i,\hat{X}_j]=i\alpha\epsilon_{ijk}\hat{X}_k.
\]
The magnitude of fuzzy sphere is specif\/ied by the dimension of  corresponding $SU(2)$ irreducible representation.
A convenient way to deal with the irreducible representation is to adopt the Schwinger boson formalism, in which the $SU(2)$ operators are given by
\[
\hat{X}_i=\frac{\alpha}{2}
\hat{\phi}^{\dagger}
\sigma_i
\hat{\phi}.
\]
Here, $\hat{\phi}=(\hat{\phi}_1,\hat{\phi}_2)^t$ stands for a Schwinger boson operator that satisfy
\[
[\hat{\phi}_{\alpha},\hat{\phi}_{\beta}^{\dagger}]=\delta_{\alpha\beta},
\]
with $\alpha,\beta=1,2$.
Square of the radius of a fuzzy two-sphere reads as
\begin{gather}
\hat{X}_i\hat{X}_i=\frac{\alpha^2}{4}\big(\hat{\phi}^{\dagger}\hat{\phi})(\hat{\phi}^{\dagger}\hat{\phi}+2\big).
\label{xsquarequantum}
\end{gather}
The commutative expression (\ref{xsquareclass}) and the fuzzy expression (\ref{xsquarequantum}) merely dif\/fer by the ``ground state energy''.
Thus, the radius of the fuzzy sphere is
\[
R_I=\frac{\alpha}{2} \sqrt{I(I+2)},
\]
where $I$ is the integer eigenvalue of the number operator  $\hat{I}\equiv\hat{\phi}^{\dagger}\hat{\phi}=\hat{\phi}_1^{\dagger}\hat{\phi}_1+\hat{\phi}_2^{\dagger}\hat{\phi}_2$. In the classical limit $I\rightarrow \infty$, the eigenvalue reproduces the radius of  commutative sphere
\[
R_I=\frac{\alpha}{2} \sqrt{I(I+2)}\rightarrow \frac{\alpha}{2}I=R.
\]
The corresponding $SU(2)$ irreducible representation  is constructed as
\begin{gather}
|m_1,m_2\rangle =\frac{1}{\sqrt{m_1! m_2!}} \hat{\phi}_1^{m_1}\hat{\phi}_2^{m_2}|0\rangle,
\label{statesonfuzzysphere}
\end{gather}
where $m_1+m_2=I$ ($m_1,m_2\ge 0$), and the degeneracy is given by $d(I)=I+1$. Apparently, there is one-to-one correspondence between the LLL monopole harmonics (\ref{LLLbasistwosphere})  and the states on  fuzzy sphere (\ref{statesonfuzzysphere}).
 Their ``dif\/ference'' is superf\/icial, coming from the corresponding representations:  Schwinger boson representation for fuzzy two-sphere,
while the $SU(2)$ coherent representation for LLL physics.
This is the basic observation of equivalence between  fuzzy geometry and  LLL physics.

\section{Graded Hopf map and fuzzy supersphere}\label{sectsuperhopf}

Before proceeding to the 2nd Hopf map,
in this section, we discuss how the relations between the Hopf map and fuzzy sphere are generalized with introducing the Grassmann numbers, mainly based on Hasebe and Kimura~\cite{hep-th/0409230}.
 The Grassmann numbers, $\eta_a$, are anticommuting numbers~\cite{Grassmann1844}
\begin{gather}
\eta_a \eta_b=-\eta_b\eta_a.
\label{algebraofgrassmann}
\end{gather}
In particular, $\eta_{a}^2=0$ (no sum for $a$).
With Grassmann numbers,
the graded Hopf map was introduced by Landi et al.~\cite{LandiMarmo1987,Bartocci1990,math-ph/9907020},
\begin{gather}
S^{3|2}    \overset{S^{1}}\longrightarrow  S^{2|2},
\label{gradedHopf}
\end{gather}
where the number on the left hand side of the slash stands for the number of bosonic (Grassmann even) coordinates and the right hand side of the slash does for the number of fermionic (Grassmann odd) coordinates.
For instance,
$S^{2|2}$ signif\/ies a supersphere with two bosonic and two fermionic coordinates.
The bosonic part of the graded Hopf map (\ref{gradedHopf}) is equivalent to the 1st Hopf map.

\subsection{Graded Hopf map and supermonopole}

As the 1st Hopf map was realized by sandwiching Pauli matrices between  two-component normalized spinors,
the graded Hopf map is realized by doing $UOSp(1|2)$ matrices between  three-component (super)spinors.
First, let us begin with the introduction of basic properties of $UOSp(1|2)$ algebra  \cite{PaisRittenberg,ScheunertNahmRittenberg,Marcu1980}.
The $UOSp(1|2)$ algebra consists of three bosonic generators~$L_i$ ($i=1,2,3$) and two fermionic generators $L_{\alpha}$ ($\alpha=\theta_1,\theta_2$) that satisfy
\begin{gather}
[L_i,L_j]=i\epsilon_{ijk}L_k, \qquad [L_i,L_{\alpha}]=\frac{1}{2}(\sigma_i)_{\beta\alpha}L_{\beta},\qquad \{L_{\alpha},L_{\beta}\}=\frac{1}{2}(\epsilon \sigma_i)_{\alpha\beta}L_i,
\label{usp12algebras}
\end{gather}
where $\epsilon=i\sigma_2$.
As realized in the f\/irst algebra of (\ref{usp12algebras}), the $UOSp(1|2)$ algebra contains the $SU(2)$
 as its subalgebra. $L_i$ transforms as   $SU(2)$ vector and $L_{\alpha}$ does as   $SU(2)$ spinor.
 The Casimir operator for   $UOSp(1|2)$ is constructed as
\[
 C=L_i L_i+\epsilon_{\alpha\beta}L_{\alpha}L_{\beta},
\]
 and its eigenvalues are given by $L(L+\frac{1}{2})$ with $L=0,1/2,1,3/2,2, \dots$.
The dimension of the corresponding irreducible representation is $4L+1$.
 The fundamental representation matrix  is given by the following $3 \times 3$ matrices
\[
l_i=\frac{1}{2}
\begin{pmatrix}
\sigma_i & 0 \\
0 & 0
\end{pmatrix},\qquad
l_{\alpha}=
\frac{1}{2}
\begin{pmatrix}
0 & \tau_{\alpha}\\
-(\epsilon\tau_{\alpha})^t & 0
\end{pmatrix},
\]
where $\sigma_i$ are the Pauli matrices, $\tau_{1}=(1,0)^t$, and $\tau_2=(0,1)^t$.
With $l_i$ and $l_{\alpha}$, the graded Hopf map~(\ref{gradedHopf}) is realized as
\begin{gather}
\varphi  \ \rightarrow \  x_i=2\varphi^{\ddagger}l_i\varphi,
\qquad \theta_{\alpha}=2\varphi^{\ddagger}l_{\alpha}\varphi,
\label{superhopf1}
\end{gather}
where $\varphi$ is a normalized three-component superspinor which we call the Hopf superspinor:
\begin{gather}
\varphi=
\begin{pmatrix}
\varphi_1 \\
\varphi_2\\
\eta
\end{pmatrix}.
\label{varphisuperhopf}
\end{gather}
Here, the f\/irst two components $\varphi_1$ and $\varphi_2$ are Grassmann even quantities, while the last component $\eta$ is Grassmann odd quantity.
The superadjoint $\ddagger$ is def\/ined as\footnote{The symbol $*$ represents the pseudo-conjugation that acts as  $(\eta^*)^*=-\eta$, $(\eta_1\eta_2)^*=\eta_1^*\eta_2^*$ for Grassmann odd quantities $\eta_1$ and $\eta_2$. See~\cite{Frappatbook} for more details.}
\[
\varphi^{\ddagger}=(\varphi_1^*,\varphi_2^*,-\eta^*),
\]
and  $\varphi$ represents  coordinates on $S^{3|2}$,  subject to the normalization condition
\[
\varphi^{\ddagger}\varphi=\varphi_1^*\varphi_1+\varphi_2^*\varphi_2-\eta^*\eta=1.
\]
$x_i$ and $\theta_{\alpha}$ given by (\ref{superhopf1}) are coordinates on  supersphere  $S^{2|2}$, since the def\/inition of  supersphere is satisf\/ied:
\begin{gather*}
x_ix_i+\epsilon_{\alpha\beta}\theta_{\alpha}\theta_{\beta}=(\varphi^{\ddagger}\varphi)^2=1.
%\label{conditiontwosphere}
\end{gather*}
Except for the south-pole of  $S^{2|2}$, the Hopf superspinor has an analytic form
\[
\varphi=\frac{1}{\sqrt{2(1+x_3)}}
\begin{pmatrix}
\displaystyle (1+x_3) \left(1-\frac{1}{4(1+x_3)}\theta \epsilon\theta\right)\vspace{1mm}\\
\displaystyle  (x_1+ix_2)\left(1+\frac{1}{4(1+x_3)}\theta \epsilon\theta\right)\vspace{1mm}\\
(1+x_3)\theta_1+(x_1+ix_2)\theta_2
\end{pmatrix}.%\label{varphiexpli}
\]
The corresponding connection is derived as
\[
A=-i\varphi^{\ddagger}d\varphi=dx_i A_i+d\theta_{\alpha}A_{\alpha},
\]
with
%%%%%%%%%%%%%%%%%%%%%%%%%%%
\begin{gather}
 A_i=-\frac{1}{2(1+x_3)}\epsilon_{ij3}x_j\biggl(1+\frac{2+x_3}{2(1+x_3)}\theta \epsilon\theta\biggr),\qquad
A_{\alpha}=-\frac{1}{2}i(x_i \sigma_i \epsilon \theta)_{\alpha}.
\label{1superconneI}
\end{gather}
 $A_i$ and $A_{\alpha}$ are   the super gauge f\/ield of  supermonopole.
The f\/ield strength is evaluated by the formula
\[
F=dA =\frac{1}{2}dx_i \wedge dx_j F_{ij}+dx_i \wedge d\theta_{\alpha} F_{i\alpha}-\frac{1}{2}d\theta_{\alpha}\wedge d\theta_{\beta} F_{\alpha\beta},
\]
where
%\[
%F_{ij}=\partial_i A_j-\partial_j A_i, \qquad F_{i\alpha}=\partial_i A_{\alpha}-\partial_{\alpha}A_i, \qquad %F_{\alpha\beta}=\partial_{\alpha}A_{\beta}+\partial_{\beta}A_{\alpha}.
%\]
%They are given by
%%%%%%%%%%%%%%%%%%%%%%%%%%%
\begin{gather}
F_{ij}=\partial_i A_j-\partial_j A_i=\frac{1}{2}\epsilon_{ij3}x_j\left(1+\frac{3}{2}\theta \epsilon\theta\right),\nonumber\\
F_{i\alpha}=\partial_i A_{\alpha}-\partial_{\alpha}A_i=-i\frac{1}{2}(\theta\sigma_j \epsilon)_{\alpha}(\delta_{ij}-3x_ix_j),\nonumber\\
F_{\alpha\beta}=\partial_{\alpha}A_{\beta}+\partial_{\beta}A_{\alpha}=-ix_i(\sigma_i \epsilon)_{\alpha\beta}\left(1+\frac{3}{2}\theta \epsilon\theta\right).
\label{superfieldstrength}
\end{gather}
%%%%%%%%%%%%%%%%%%%%%%%%%%%%%%%
The analytic form of the Hopf superspinor except for the north pole is given by
\[
\varphi'
=\frac{1}{\sqrt{2(1-x_3)}}
\begin{pmatrix}
\displaystyle  (x_1-ix_2) \left(1+\frac{1}{4(1-x_3)}\theta \epsilon\theta\right)\vspace{1mm}\\
\displaystyle  (1-x_3)\left(1-\frac{1}{4(1-x_3)}\theta \epsilon\theta\right)\vspace{1mm}\\
(x_1-ix_2)\theta_1+(1-x_3)\theta_2
\end{pmatrix}, %\label{varphidashexpli}
\]
and the corresponding gauge f\/ield is obtained as
\begin{gather}
A'_i=\frac{1}{2(1-x_3)}\epsilon_{ij3}x_j\left(1+\frac{2-x_3}{2(1-x_3)}\theta \epsilon\theta\right),\qquad
A'_{\alpha}=-i\frac{1}{2}(x_i\sigma_i \epsilon \theta)_{\alpha}=A_{\alpha}.
\label{supergaugefielddiffgauge}
\end{gather}
The f\/ield strength $F'=dA'$ is same as (\ref{superfieldstrength}), suggesting
the two expressions of the Hopf superspinor are related by the transformation
\[
\varphi'=\varphi \cdot g=g \cdot \varphi.
\]
Here, $g$ denotes $\mathcal{U}(1)$ gauge group element given by
\[
g=e^{-i \chi}=\frac{x_1-ix_2}{\sqrt{1-x_3^2}}\left(1+\frac{1}{2(1-x_3^2)}\theta \epsilon\theta\right),
\]
with $\chi$  given by $\tan\chi=\frac{x_2}{x_1}$.
The $\mathcal{U}(1)$ group element yields
\[
-ig^*dg=\frac{1}{1-x_3^2}\epsilon_{ij3}x_i dx_j \left(1+\frac{1}{1-x_3^2}\theta \epsilon\theta\right),
\]
and  $A$ (\ref{1superconneI}) and $A'$ (\ref{supergaugefielddiffgauge}) are expressed as
\begin{gather*}
A= i\frac{1}{2}\left(1-x_3(1+\frac{1}{2}\theta \epsilon\theta)\right)g^*dg+d\theta_{\alpha}A_{\alpha},\nonumber\\
A'=-i\frac{1}{2}\left(1+x_3(1+\frac{1}{2}\theta \epsilon\theta)\right)g^*dg+d\theta_{\alpha}A_{\alpha}.
\end{gather*}
Then, obviously  the gauge f\/ields are related as
\[
A'=A-ig^*dg,
\]
and then
\[
F'=F.
\]

\subsection[$UOSp(1|2)$ Landau model]{$\boldsymbol{UOSp(1|2)}$ Landau model}

Next, we consider  Landau problem on a supersphere in supermonopole background~\cite{hep-th/0409230}.
The supermonopole gauge f\/ield is given by
\[
A_i=-\frac{I}{2r(r+x_3)}\epsilon_{ij3}x_j\left(1+\frac{2r+x_3}{2r^2(r+x_3)}\theta \epsilon\theta\right),\qquad
A_{\alpha}=-\frac{I}{2r^3}i(x_i \sigma_i \epsilon \theta)_{\alpha},
%\label{1superconneI2}
\]
where $\frac{I}{2}$ ($I$ denotes an integer) is a magnetic charge of the supermonopole.
Generalizing the $SO(3)$ Landau Hamiltonian (\ref{on2DsphereLandau}) to  $UOSp(1|2)$ form,
 we have
\begin{gather}
H=\frac{1}{2MR^2}(\Lambda_i^2+ \epsilon_{\alpha\beta}\Lambda_{\alpha}\Lambda_{\beta}),
\label{superLandau}
\end{gather}
where $\Lambda_i$ and $\Lambda_{\alpha}$ are the bosonic and fermionic components of covariant angular momentum:
\begin{gather*}
 \Lambda_i=-i\epsilon_{ijk}x_jD_k+\frac{1}{2}\theta_{\alpha}(\sigma_i)_{\alpha\beta}D_{\beta},\qquad
 \Lambda_{\alpha}=\frac{1}{2}( \epsilon\sigma_i)_{\alpha\beta}x_iD_{\beta}-\frac{1}{2}\theta_{\beta}(\sigma_i)_{\beta\alpha}D_i,
\end{gather*}
with $D_i=\partial_i+iA_i$ and $D_{\alpha}=\partial_{\alpha}+iA_{\alpha}$.
$\Lambda_i$ and $\Lambda_{\alpha}$ obey the following graded commutation relations
\begin{gather*}
[\Lambda_i,\Lambda_j]=i\epsilon_{ijk}(\Lambda_k-r^2F_k), \qquad [\Lambda_i,\Lambda_{\alpha}]=\frac{1}{2}(\sigma_i)_{\beta\alpha}(\Lambda_{\beta}-r^2F_{\beta}),\nonumber\\
 \{\Lambda_{\alpha},\Lambda_{\beta}\}=\frac{1}{2}( \epsilon\sigma_i)_{\alpha\beta}(\Lambda_i-r^2F_i),
%\label{algebrasofcovariantangular}
\end{gather*}
where  $F_i$  and $F_{\alpha}$ are
\[
F_i=\frac{I}{2r^3}x_i,\qquad F_{\alpha}=\frac{I}{2r^3}\theta_{\alpha}.
\]
Just as in the case of the $SO(3)$ Landau model,  the covariant angular momentum is not a conserved quantity, in the sense $[\Lambda_i,H]\neq 0$, $[\Lambda_{\alpha},H]\neq 0$.
Conserved angular momentum is constructed as
\[
L_i=\Lambda_i+r^2F_i,\qquad L_{\alpha}=\Lambda_{\alpha}+r^2F_{\alpha}.
\]
It is straightforward to see $L_i$ and $L_{\alpha}$ satisfy the $UOSp(1|2)$ algebra
(\ref{usp12algebras}).
%%%%%%%%%%%%%%%%%%%%%%%%%
Since the covariant angular momentum and the f\/ield strength are orthogonal in the supersymmetric sense: $\Lambda_i F_i+ \epsilon_{\alpha\beta}\Lambda_{\alpha}F_{\beta}= F_i\Lambda_i+ \epsilon_{\alpha\beta}F_{\alpha}\Lambda_{\beta}=0$,
the $UOSp(1|2)$ Landau Hamiltonian (\ref{superLandau}) can be rewritten  as
\[
H=\frac{1}{2MR^2}\left(L_i^2+ \epsilon_{\alpha\beta}L_{\alpha}L_{\beta}-\frac{I^2}{4}\right).
\]
With the Casimir index $J=\frac{I}{2}+n$,
the energy eigenvalue is derived as
\[
E_n=\frac{1}{2MR^2}\left(J\left(J+\frac{1}{2}\right)-\frac{I^2}{4}\right)\Big|_{J=n+\frac{I}{2}}
=\frac{1}{2MR^2}\left(n\left(n+\frac{1}{2}\right)+I\left(n+\frac{I}{4}\right)\right),
\]
and the degeneracy in the $n$th Landau level is
\[
d_n=4J+1|_{J=n+\frac{I}{2}}=4n+2I+1.
\]
In particular in the LLL ($n=0$), the degeneracy becomes
\begin{gather}
d_{\rm LLL}=2I+1.
\label{LLLdegeneracy}
\end{gather}
The eigenstates of the $UOSp(1|2)$ Hamiltonian are referred to
the supermonopole harmonics, and the LLL eigenstates are constructed by taking  symmetric products of the components of the Hopf superspinor~(\ref{varphisuperhopf}):
%%%%%%%%%%%%%%%%%%%
\begin{gather}
\varphi^{B~(m_1,m_2)}_{\rm LLL}=\sqrt{\frac{I!}{m_1! m_2 !}}\varphi_1^{m_1}\varphi_2^{m_2},\qquad
\varphi^{F~(n_1,n_2)}_{\rm LLL}=\sqrt{\frac{I!}{n_1 ! n_2!}}\varphi_1^{n_1}\varphi_2^{n_2}\eta,
\label{supermonopoleharmonicsLLL}
\end{gather}
where $m_1+m_2=n_1+n_2+1=I$ $(m_1,m_2,n_1,n_2\ge 0)$. The total number of $\varphi_{\rm LLL}^{B~(m_1,m_2)}$ and  $\varphi_{\rm LLL}^{F~(n_1,n_2)}$ is $(I+1)+(I)=2I+1$, which coincides with~(\ref{LLLdegeneracy}).
In the LLL, the $UOSp(1|2)$ angular momentum is reduced to
\[
L_i \ \rightarrow \ R^2F_i=\frac{I}{2R}x_i,\qquad L_{\alpha} \ \rightarrow \ R^2F_{\alpha}=\frac{I}{2R}\theta_{\alpha},
\]
and coordinates on  supersphere are identif\/ied with the $UOSp(1|2)$ operators
\[
X_i={\alpha}L_i,\qquad \Theta_{\alpha}=\alpha L_{\alpha},
\]
which satisfy the algebra def\/ining  fuzzy supersphere:
\begin{gather}
[X_i,X_j]=i\alpha\epsilon_{ijk}X_k, \qquad [X_i,\Theta_{\alpha}]=\frac{\alpha}{2}(\sigma_i)_{\beta\alpha}\Theta_{\beta},
\qquad \{\Theta_{\alpha},\Theta_{\beta}\}=\frac{\alpha}{2}( \epsilon\sigma_i)_{\alpha\beta}X_i.
\label{fuzzysupersphereinLLL}
\end{gather}

With Lagrange formalism, we reconsider the LLL physics.
The present one-particle Lagrangian on a~supersphere is given by
\[
L=\frac{M}{2}\big(\dot{x}_i^2+ \epsilon_{\alpha\beta}\dot{\theta}_{\alpha}\dot{\theta}_{\beta}\big)+A_i\dot{x}_i+\dot{\theta}_{\alpha}A_{\alpha}
\]
with the constraint
\[
x_i x_i+ \epsilon_{\alpha\beta}\theta_{\alpha}\theta_{\beta}=R^2.
\]
In the LLL, the kinetic term is quenched, and the Lagrangian takes the form of
\begin{gather}
L_{\rm LLL}=\dot{x}_i A_i +\dot{\theta}_{\alpha}A_{\alpha}=-iI\varphi^{\ddagger}\frac{d}{dt}\varphi,
\label{superLLLlag}
\end{gather}
with the constraint
\begin{gather}
\varphi^{\ddagger}\varphi=1.
\label{normalizationvarphi}
\end{gather}
From the LLL Lagrangian (\ref{superLLLlag}), the canonical momentum of $\varphi$ is derived as $\pi=iI{\varphi}^*$, and from the commutation relation between $\varphi$ and $\pi$, the complex conjugation is quantized as
\begin{gather}
\varphi^*=\frac{1}{I}\frac{\partial}{\partial\varphi}.
\label{varphideriv}
\end{gather}
After quantization, the normalization condition (\ref{normalizationvarphi}) is imposed on the LLL basis:
\[
\varphi^t \frac{\partial}{\partial \varphi}\varphi_{\rm LLL}=I\varphi_{\rm LLL}.
\]
One may conf\/irm the LLL basis (\ref{supermonopoleharmonicsLLL}) satisf\/ies the condition.
By inserting (\ref{varphideriv}) to the graded Hopf map (\ref{superhopf1}), we obtain
\[
X_i=\alpha\varphi^t l_i\frac{\partial}{\partial\varphi},
\qquad \Theta_{\alpha}=\alpha\varphi^t l_{\alpha}\frac{\partial}{\partial\varphi}.
\]
Apparently, they satisfy the algebra of  fuzzy supersphere (\ref{fuzzysupersphereinLLL}).
Thus, in the case of the fuzzy two-sphere, the appearance of fuzzy supersphere  in LLL is naturally understood in the context of  the graded Hopf map.

\subsection{Fuzzy supersphere}

Fuzzy supersphere is constructed by taking symmetric representation of the $UOSp(1|2)$ group \cite{hep-th/9507074,math-ph/9804013,hep-th/0204170,hep-th/0511114}.
We f\/irst introduce a superspinor extension of the Schwinger operator
\[
\hat{\varphi}=
\begin{pmatrix}
\hat{\varphi}_1\\
\hat{\varphi}_2\\
\hat{\eta}
\end{pmatrix},
\]
where $\hat{\varphi_1}$ and $\hat{\varphi}_2$ are Schwinger boson operators, and  $\hat{\eta}$ is a fermion operator:  $[\hat{\varphi}_i,\hat{\varphi}_j^{\dagger}]=\delta_{ij}$,  $[\hat{\varphi}_i,\hat{\eta}]=0$ and  $\{\hat{\eta},\hat{\eta}^{\dagger}\}=1$.
With such  Schwinger superoperator,   coordinates on fuzzy supersphere are constructed as
\[
\hat{X}_i= \alpha
\hat{\varphi}^{\dagger}
l_i
\hat{\varphi},\qquad
\hat{\Theta}_{\alpha}=
\alpha
\hat{\varphi}^{\dagger}
l_{\alpha}
\hat{\varphi}.
\]
Square of the radius of fuzzy supersphere is given by
\[
\hat{X}_i \hat{X}_i+ \epsilon_{\alpha\beta}\hat{\Theta}_{\alpha}\hat{\Theta}_{\beta}
=\frac{\alpha^2}{4}(\hat{\varphi}^{\dagger}\hat{\varphi})(\hat{\varphi}^{\dagger}\hat{\varphi}+1),
\]
and then the radius is specif\/ied by
the integer eigenvalue $I$ of the number operator $\hat{I}\equiv\hat{\varphi}^{\dagger}\hat{\varphi}=\hat{\varphi}^{\dagger}_1\hat{\varphi}_1+ \hat{\varphi}^{\dagger}_2\hat{\varphi}_2+\hat{\eta}^{\dagger}\hat{\eta}$ as
\[
R_I=\frac{\alpha}{2}\sqrt{I(I+1)}.
\]
The $UOSp(1|2)$ symmetric irreducible representation is explicitly constructed as
\begin{gather*}
 |m_1,m_2\rangle=\frac{1}{\sqrt{m_1! m_2!}}(\hat{\varphi}_1^{\dagger})^{m_1}(\hat{\varphi}_2^{\dagger})^{m_2}|0\rangle,\qquad
|n_1,n_2\rangle =\frac{1}{\sqrt{n_1! n_2!}}(\hat{\varphi}_1^{\dagger})^{n_1}(\hat{\varphi}_2^{\dagger})^{n_2} \hat{\eta}^{\dagger}|0\rangle,
\end{gather*}
with $m_1+m_2=n_1+n_2+1=I$ ($m_1,m_2,n_1,n_2\ge 0$). Thus, the total number of  the states constructing fuzzy supersphere is $d=(I+1)+(I)=2I+1$.
There is one-to-one correspondence between the supermonopole harmonics in LLL  and the states on fuzzy supersphere. Especially, the Hopf superspinor corresponds to the superspin coherent state of Schwinger superoperator.

\section{2nd Hopf map and fuzzy four-sphere}\label{Sec2ndHopf}

In this section, we discuss relations between the 2nd Hopf map
\[
S^7 \overset{S^3}\longrightarrow S^4
%\label{2ndcopacthopf}
\]
 and  fuzzy four-sphere.
Though the 2nd and 3rd Hopf maps were f\/irst introduced by Hopf~\cite{Hopf1935}, we follow the realization given by Zhang and Hu \cite{Zhang2001}
in the following discussions.

\subsection[2nd Hopf map and $SU(2)$ monopole]{2nd Hopf map and $\boldsymbol{SU(2)}$ monopole}

Realization of the 2nd Hopf map is easily performed by replacing the imaginary unit with the imaginary quaternions, $q_1$, $q_2$, $q_3$.  The Pauli matrices (\ref{paulimatrices}) are  promoted to the following quaternionic Pauli matrices
\begin{gather}
\gamma_1=
\begin{pmatrix}
0 & -q_1\\
q_1 & 0
\end{pmatrix},\qquad \gamma_2=
\begin{pmatrix}
0 & -q_2\\
q_2 & 0
\end{pmatrix}, \qquad \gamma_3=
\begin{pmatrix}
0 & -q_3\\
q_3 & 0
\end{pmatrix},\nonumber\\
\gamma_4=
\begin{pmatrix}
0 & 1\\
1 & 0
\end{pmatrix}, \qquad \gamma_5=
\begin{pmatrix}
1 & 0\\
0 & -1
\end{pmatrix}.\label{quaternionicpauli}
\end{gather}
$\gamma_i$ $(i=1,2,3)$ correspond to $\sigma_2$, $\gamma_4$ to $\sigma_1$, and $\gamma_5$ to $\sigma_3$, respectively.
With such quaternionic ``Pauli matrices'', the 2nd Hopf map is realized as
\begin{gather}
\psi \ \to \ \psi^{\dagger}\gamma_a\psi=x_a,
\label{quaternionicHopf}
\end{gather}
where $a=1,2, \dots,5$, and $\psi$ is a two-component {\it quaternionic} (quaternion-valued) spinor satisfying the normalization condition $\psi^{\dagger}\psi=1$.
As in the 1st Hopf map,  $x_a$ given by the map~(\ref{quaternionicHopf}) automatically satisf\/ies the condition $x_a x_a=(\psi^{\dagger}\psi)^2=1$.
Like the 1st Hopf spinor,  the quaternionic Hopf spinor has an analytic form, except for the south pole, as
\[
\psi=
\frac{1}{\sqrt{2(1+x_5)}}
\begin{pmatrix}
1+x_5 \\
x_4+q_i x_i
\end{pmatrix},
\]
and, except for the north pole, as
\[
\psi'=\frac{1}{\sqrt{2(1-x_5)}}
\begin{pmatrix}
x_4-q_ix_i\\
1-x_5
\end{pmatrix},
\]
where $i=1,2,3$ are summed over.
These two expressions are related by the transformation
\[
\psi'=\psi \cdot g=g \cdot \psi,
\]
where $g$ is a quaternionic $U(1)$ group element given by
\[
g=e^{-q_i \chi_i}= \cos ( \chi)-q_i\frac{ \chi_i}{\chi}\sin ( \chi)=\frac{1}{\sqrt{1-x_5^2}}(x_4-q_ix_i).
\]
Here, $\chi_i$ and its magnitude  $\chi=\sqrt{ \chi_i^2}$ are determined by  $\frac{ \chi_i}{\chi}\tan( \chi)=\frac{1}{x_4}x_i$.
It is possible to pursue the discussions with use of quaternions, but for later convenience, we utilize the Pauli matrix representation of the imaginary quaternions:
\[
q_1=-i\sigma_1,\qquad q_2=-i\sigma_2,\qquad q_3=-i\sigma_3.
\]
The quaternionic $U(1)$ group is usually denoted as $U(1,\mathbb{H})$, and as obvious from the above identif\/ication, $U(1,\mathbb{H})$ is isomorphic to $SU(2)$.
The quaternionic Pauli matrices (\ref{quaternionicpauli}) are now  represented by the following   $SO(5)$ gamma matrices:
\begin{gather*}
\gamma_1=
\begin{pmatrix}
0 & i\sigma_1 \\
-i\sigma_1 & 0
\end{pmatrix}, \qquad \gamma_2=
\begin{pmatrix}
0 & i\sigma_2 \\
-i\sigma_2 & 0
\end{pmatrix}, \qquad \gamma_3=
\begin{pmatrix}
0 & i\sigma_3 \\
-i\sigma_3 & 0
\end{pmatrix}, \\
\gamma_4=
\begin{pmatrix}
0 & 1 \\
1 & 0
\end{pmatrix}, \qquad \gamma_5=
\begin{pmatrix}
1 & 0 \\
0 & -1
\end{pmatrix},
\end{gather*}
which satisfy $\{\gamma_a,\gamma_b\}=2\delta_{ab}$ ($a,b=1,2,3,4,5$).
Corresponding to the quaternionic Hopf spinor,
we introduce a $SO(5)$ four-component spinor (the 2nd Hopf spinor)
\begin{gather}
\psi=
\begin{pmatrix}
\psi_1\\
\psi_2 \\
\psi_3\\
\psi_4
\end{pmatrix},
\label{fourcomphopfspi}
\end{gather}
subject to the constraint
\[
\psi^{\dagger}\psi=1.
\]
With  $\psi$, the 2nd Hopf map (\ref{quaternionicHopf}) is rephrased as
\begin{gather}
\psi \rightarrow x_a=\psi^{\dagger}\gamma_a \psi.
\label{quaternionichopf2}
\end{gather}
It is easy to check that $x_a$  satisf\/ies the condition of $S^4$:
\[
x_a x_a=(\psi^{\dagger}\psi)^2=1.
\]
An analytic form of the 2nd Hopf spinor, except for the south pole, is given by
\begin{gather}
\psi=\frac{1}{\sqrt{2(1+x_5)}}
\begin{pmatrix}
(1+x_5)\phi \\
(x_4-i\sigma_i x_i)\phi
\end{pmatrix},
\label{explicitnorth2ndhopf}
\end{gather}
where $\phi$ is the 1st Hopf spinor representing $S^3$-f\/ibre.
The connection is derived as
\begin{gather}
A=-i\psi^{\dagger}d\psi=\phi^{\dagger} dx_a A_a \phi
\label{psivariationSU2}
\end{gather}
where
\begin{gather}
A_{\mu}=-\frac{1}{2(1+x_5)}\eta_{\mu\nu i}^{+}  x_{\nu} \sigma_i,\qquad A_5=0.
\label{su2gaugeori}
\end{gather}
They are the $SU(2)$ gauge f\/ield of Yang monopole~\cite{Yang1978}.
Here, $\eta_{\mu\nu i}^+$ signif\/ies the 'tHooft eta-symbol of instanton~\cite{tHooft1976}:
\[
\eta_{\mu\nu i}^{+}=\epsilon_{\mu\nu i4}+\delta_{\mu i}\delta_{\nu 4}-\delta_{\mu 4}\delta_{\nu i}.
\]
The corresponding f\/ield strength $F=dA +i A\wedge A=\frac{1}{2}dx_a\wedge dx_b F_{ab}$:
\[
F_{ab}=\partial_a A_b-\partial_b A_a +i[A_a,A_b],
\]
is evaluated as
%%%%%%%%%%%%%%%%%%%%%%%%%%%%
\begin{gather}
 F_{\mu\nu}=-x_{\mu}A_{\nu}+x_{\nu}A_{\mu}+\frac{1}{2}\eta_{\mu\nu i}^{+}\sigma_i,\qquad
 F_{\mu 5}=-F_{5\mu}=(1+x_5)A_{\mu}.\label{SU(2)strength}
\end{gather}
Another analytic form of the 2nd Hopf spinor, except for the north pole, is given by
\[
\psi'=\frac{1}{\sqrt{2(1-x_5)}}
\begin{pmatrix}
(x_4+ix_i\sigma_i)\phi \\
(1-x_5)\phi
\end{pmatrix}.
\]
The corresponding connection is calculated as
\begin{gather}
A'_{\mu}=-\frac{1}{2(1-x_5)}\eta_{\mu\nu i}^{-}x_{\nu}\sigma_i,\qquad A'_5=0,
\label{su2gaugediff}
\end{gather}
where
\[
\eta_{\mu\nu i}^{-}=\epsilon_{\mu\nu i4}-\delta_{\mu i}\delta_{\nu 4}+\delta_{\mu 4}\delta_{\nu i},
\]
and the f\/ield strength is
%%%%%%%%%%%%%%%%%%%%%%%%%%%%
\begin{gather*}
 F'_{\mu\nu}=x_{\mu}A'_{\nu}-x_{\nu}A'_{\mu}+\frac{1}{2}\eta_{\mu\nu i}^{-}\sigma_i,\qquad
 F'_{\mu 5}=-F'_{5\mu}=(1-x_5)A'_{\mu}.%\label{SU(2)strengthdash}
\end{gather*}
As is well known, the 'tHooft eta-symbol satisf\/ies the self (anti-self) dual relation
\[
\eta_{\mu\nu i}^{\pm}=\pm\frac{1}{2}\epsilon_{\mu\nu\rho\sigma}\eta_{\rho\sigma i}^{\pm}.
\]
The two expressions,  $\psi$ and $\psi'$, are related by the $SU(2)$ transition function
\[
g=\frac{1}{\sqrt{1-{x_5}^2}}(x_4+ix_i\sigma_i),
\]
which yields
\[
 -ig^{\dagger}dg=-\frac{1}{1-{x_5}^2}\eta_{\mu\nu i}^{-}x_{\nu}\sigma_{i}dx_{\mu},\qquad
-idg g^{\dagger}=\frac{1}{1-{x_5}^2}\eta_{\mu\nu i}^{+}x_{\nu}\sigma_idx_{\mu},
\]
and the gauge f\/ields (\ref{su2gaugeori}) and (\ref{su2gaugediff}) are concisely represented as
\[
 A=\frac{1-x_5}{2}idg g^{\dagger},\qquad
A'=-\frac{1+x_5}{2}ig^{\dagger}dg.
\]
Then,  $A$ and $A'$ are related as
\[
A'=g^{\dagger}A g-ig^{\dagger}dg,
\]
and their f\/ield strengths are also
\[
F'=g^{\dagger}F g.
\]
This manifests the non-trivial topology of the $SU(2)$ bundle on a four-sphere.
In the Language of the homotopy theorem, the non-trivial topology of the $SU(2)$ bundle is expressed by
\[
\pi_{3}(SU(2))\simeq \mathbb{Z},
\]
which is specif\/ied by the 2nd Chern number
\[
c_2=\frac{1}{2(2\pi)^2}\int_{S^4} {\rm tr}\, (F\wedge F).
\]

\subsection[$SO(5)$ Landau model]{$\boldsymbol{SO(5)}$ Landau model}

We next explore the Landau problem in 5D space~\cite{Zhang2001}. The Landau Hamiltonian is given by
\begin{gather}
H=-\frac{1}{2N}D_a^2=-\frac{1}{2M}\frac{\partial^2}{\partial r^2}-\frac{2}{Mr}\frac{\partial}{\partial r}+\frac{1}{2Mr^2}\sum_{a<b}\Lambda_{ab}^2,
\label{5Dlandauhamilto}
\end{gather}
where $D_a=\partial_a +iA_a$, $r=\sqrt{x_a x_a}$, and $\Lambda_{ab}$ are the $SO(5)$ covariant angular momentum
\[
\Lambda_{ab}=-ix_aD_b+ix_bD_a.
\]
As in the previous 3D case,  $\Lambda_{ab}$ do not satisfy a closed algebra, but satisfy
\begin{gather*}
[\Lambda_{ab},\Lambda_{cd}]= i(\delta_{ac}\Lambda_{bd}+\delta_{bd}\Lambda_{ac}-\delta_{bc}\Lambda_{ad}-\delta_{ad}\Lambda_{bc})\\
\hphantom{[\Lambda_{ab},\Lambda_{cd}]=}{}
-i(x_a x_c F_{bd}+x_{b}x_{d}F_{ac}-x_{b}x_{c}F_{ad}-x_{a}x_{d}F_{bc} ),
\end{gather*}
where $F_{ab}$ are given by (\ref{SU(2)strength}).
The $SO(5)$ conserved angular momentum is constructed as
\[
L_{ab}=\Lambda_{ab}+r^2F_{ab}.
\]
On a four-sphere, the Hamiltonian (\ref{5Dlandauhamilto}) is reduced to the $SO(5)$ Landau Hamiltonian
\[
H=\frac{1}{2MR^2}\sum_{a<b}\Lambda_{ab}^2,
\]
which is rewritten as
\[
H=\frac{1}{2MR^2}\sum_{a<b}\big(L_{ab}^2-R^4F_{ab}^2\big)=\frac{1}{2MR^2}\sum_{a<b}\left(L_{ab}^2-\frac{1}{2}I(I+4)\right),
\]
where the orthogonality, $\Lambda_{ab}F_{ab}=F_{ab}\Lambda_{ab}=0$, was used.
Thus, the energy eigenvalue problem of the Hamiltonian is again boiled down to the problem of obtaining irreducible representation of the $SO(5)$ Casimir.
Since the $SO(5)$ group has two Casimirs and the irreducible representations are specif\/ied by two indices $[\lambda_1,\lambda_2]$.\footnote{We follow the notation in~\cite{IachelloBook}.}   With the identif\/ication $[\lambda_1,\lambda_2]=[n+\frac{I}{2},\frac{I}{2}]$,
 the energy eigenvalues are expressed as
\begin{gather}
E_n=\frac{1}{2MR^2}\big(n^2+n(I+3)+I\big).
\label{SO5energyegenvalues}
\end{gather}
The integer $I$ $(I=0,1,2, \dots)$ specif\/ies the $SU(2)$ representation of the  gauge f\/ield, while
$n$ $(n=0,1,2, \dots)$ does the Landau level.  The states in the LLL ($n=0$) correspond to the fully symmetric spinor representation of $SO(5)$,  $[\frac{I}{2},\frac{I}{2}]$.
The degeneracy in the $n$th Landau level is derived as
\begin{gather}
d_n=\frac{1}{3!}(n+1)(I+1)(n+I+2)(2n+I+3).
\label{nthdegeneracyonfoursphere}
\end{gather}
The eigenstates of the Hamiltonian are given by the $SU(2)$ monopole harmonics~\cite{CNyang1978}.
In particular, the $SU(2)$ monopole harmonics in the LLL are simply constructed by taking the symmetric products of the components of the 2nd Hopf spinor~(\ref{fourcomphopfspi})
\begin{gather}
\psi_{\rm LLL}^{(m_1,m_2,m_3,m_4)}=\sqrt{\frac{I!}{m_1! m_2 ! m_3 ! m_4 !}}\psi_1^{m_1}\psi_2^{m_2}\psi_3^{m_3}\psi_4^{m_4}
\label{SO(5)LLLbases}
\end{gather}
with $m_1+m_2+m_3+m_4=I$ ($m_1,m_2,m_3,m_4 \ge 0$).
Then, the LLL degeneracy is given by
\[
d_{\rm LLL}=\frac{1}{3!}(I+3)(I+2)(I+1),
%\label{LLLdegeonfoursphere}
\]
which actually coincides with $d_{n=0}$ of (\ref{nthdegeneracyonfoursphere}).
Up to now, everything is parallel with the $SO(3)$ Landau model, but  emergence of fuzzy four-sphere in LLL is not transparent unlike
 the fuzzy two-sphere case.

With Lagrange formalism, we revisit LLL physics of the $SO(5)$ Landau model.
The present one-particle Lagrangian on a four-sphere is given by
%%%%%%%%%%%%%%%%%%%%%%
\begin{gather}
L=\frac{M}{2}\dot{x}_a\dot{x}_a+\dot{x}_aA_a,
\label{originalSO(5)symmLag}
\end{gather}
%%%%%%%%%%%%%%%%%%%%%%
with a constraint
\begin{gather}
x_ax_a=R^2.
\label{contrs4}
\end{gather}
 In the LLL, the interaction term survives to yield
\begin{gather}
L_{\rm LLL}=\dot{x}_a A_a,
\label{su4llllagrangin}
\end{gather}
and the relation (\ref{psivariationSU2}) implies
\begin{gather}
L_{\rm LLL}=-i\psi^{\dagger}\frac{d}{dt}\psi,
\label{psiLLLLag}
\end{gather}
and also the constraint (\ref{contrs4}) is rewritten as
\begin{gather}\label{****}
\psi^{\dagger}\psi=1.
\end{gather}
Interestingly, in the LLL, the original $SO(5)$ symmetry of the Lagrangian (\ref{originalSO(5)symmLag}) is enhanced to the $SU(4)$  symmetry:  the rotational symmetry of the 2nd Hopf spinor.
We treat the 2nd spinor as the fundamental variable and apply the quantization condition.
After quantization, the complex conjugate spinor is regarded as the derivative
$\psi^*= \frac{1}{I}\frac{\partial}{\partial\psi}$, and
the normalization condition~\eqref{****} is translated to the LLL condition
\begin{gather}
\psi^t\frac{d}{d\psi} \psi_{\rm LLL}=I\psi_{\rm LLL}.
\label{condsu5lll}
\end{gather}
The LLL states (\ref{SO(5)LLLbases}) indeed satisfy the condition (\ref{condsu5lll}).
Here, we comment on the origin of the $SU(4)$ symmetry and its relation to the 2nd Hopf map.
In the LLL, the 2nd Hopf spinor plays a~primary role, and the total manifold $S^7$ naturally appears in LLL. Projecting out the $U(1)$ phase from~$S^7$, we obtain the structure of
\[
\mathbb{C}P^3\simeq S^7/S^1.
\]
This suggests physical equivalence between the LLL of $SO(5)$ Landau model and that of $SU(4)$ Landau model on $\mathbb{C}P^3$.
(Detail discussions on physical equivalence between two Lagran\-gians (\ref{su4llllagrangin}) and (\ref{psiLLLLag}) are found in~\cite{Bernevig2002effective}, and see Appendix~\ref{sectioncomplexprojective2} also.)
The appearance of $\mathbb{C}P^3$ can also be understood as follows.
 As mentioned in Introduction, $S^4$ is not a K\"ahler manifold that accommodates  symplectic structure.  The ``minimally extended'' symplectic manifold of~$S^4$ is~$\mathbb{C}P^3$, which is given by the coset $ SU(4)/U(3)$, and then  the $SU(4)$ structure naturally appears.
Such observation is completely consistent with the mathematical expression of the fuzzy four-sphere, since
\[
S_{F}^4\simeq SO(5)/U(2)\simeq SU(4)/U(3)\simeq \mathbb{C}P^3,
\]
where we used $SO(6)/SO(5)\simeq S^5\simeq U(3)/U(2)$ and $SO(6)\simeq SU(4)$.
By inserting the derivative expression of the complex 2nd Hopf spinor to the 2nd Hopf map~(\ref{quaternionichopf2}), we f\/ind that
the coordinate on $S^4$ is expressed by the following operator
\[
X_a=\alpha\psi^{t}\gamma_a \frac{\partial}{\partial\psi}.
\]
As we shall see in the next subsection, the $SU(4)$ structure also appears in the enhanced algebra of $X_a$.

\subsection{Fuzzy four-sphere}

The fuzzy four-sphere is constructed by taking a fully symmetric representation of the $SO(5)$ spinor \cite{hep-th/9602115,hep-th/9712105,Ho2002,Kimura2002,hep-th/0105006,AzumaBagnoud2003,Abe2004}.  As in the fuzzy two-sphere case,
the Schwinger boson formalism is useful to construct   coordinates on fuzzy four-sphere
\begin{gather}
\hat{X}_a=\frac{\alpha}{2}\hat{\psi}^{\dagger}
\gamma_a
\hat{\psi},
\label{so5gammaop}
\end{gather}
where $\hat{\psi}$ is a four-component Schwinger boson operator satisfying  $[\hat{\psi}_{\alpha},\hat{\psi}_{\beta}^{\dagger}]=\delta_{\alpha\beta}$  $(\alpha,\beta=1,2,3,4$).
The commutations relation of $X_a$  gives
\[
[\hat{X}_a,\hat{X}_b]=\frac{\alpha}{2}\hat{X}_{ab}
\]
where $\hat{X}_{ab}$ is the $SO(5)$ generator of the form
\begin{gather}
\hat{X}_{ab}=\frac{\alpha}{2}\hat{\psi}^{\dagger}\sigma_{ab}\hat{\psi}
\label{so5geneop}
\end{gather}
with  $\sigma_{ab}=-\frac{i}{4}[\gamma_a,\gamma_b]$.
It is important to notice, unlike the case of fuzzy two-sphere, the fuzzy coordinates do {\it not} satisfy a closed algebra by themselves  but yield the $SO(5)$ generators.
With the $SO(5)$ generators $\hat{X}_{ab}$, the fuzzy coordinates satisfy the following closed algebra,
\begin{gather*}
[\hat{X}_a,\hat{X}_b]=\frac{\alpha}{2}\hat{X}_{ab},\qquad
 [\hat{X}_a,\hat{X}_{bc}]= -i\frac{\alpha}{2}(\delta_{ab}\hat{X}_c-\delta_{ac}\hat{X}_b),\nonumber\\
 [\hat{X}_{ab},\hat{X}_{cd}]=i\frac{\alpha}{2}
(\delta_{ac}\hat{X}_{bd}-\delta_{ad}\hat{X}_{bc}+\delta_{bc}\hat{X}_{ad}-\delta_{bd}\hat{X}_{ac}).
\end{gather*}
By identifying $\hat{X}_{a6}=\frac{1}{2}\hat{X}_a$ and $\hat{X}_{ab}=\hat{X}_{ab}$, we f\/ind the above algebra is concisely expressed by the $SO(6)$ algebra,
\[
[\hat{X}_{AB},\hat{X}_{CD}]
=i\frac{\alpha}{2}(\delta_{AC}\hat{X}_{BD}-\delta_{AD}\hat{X}_{BC}+\delta_{BC}\hat{X}_{AD}-\delta_{BD}\hat{X}_{AC}),
\]
where $A,B=1,2,\dots,6$.
Thus, the algebra def\/ining fuzzy four-sphere is  $SO(6)\simeq SU(4)$.
We encountered the $SU(4)$ structure again, and the fuzzy manifold naturally def\/ined by $SU(4)$ algebra is  fuzzy $\mathbb{C}P^3$ (see Appendix~\ref{sectioncomplexprojective1}).  The enhanced $SU(4)$ algebra with extra $X_{ab}$ coordinates accounts for the existence of extra fuzzy-dimensions~\cite{Ho2002,Kimura2002}.  Interestingly,
$\mathbb{C}P^3$ is locally expressed as the two-sphere f\/ibration over the four-sphere:
\begin{gather*}
\mathbb{C}P^3\approx S^4\times S^2,
%\label{cp3fibrations2overs4}
\end{gather*}
  since $\mathbb{C}P^3\simeq S^7/S^1 \approx S^4\times S^3/S^1$.
 The ``extra dimension'' of $S_F^4$ can be understood as two-sphere f\/ibration over the four-sphere.

Square of the radius of  fuzzy four-sphere is calculated as
\begin{gather*}
\hat{X}_a\hat{X}_a=\frac{\alpha^2}{4}(\hat{\psi}^{\dagger}\hat{\psi})(\hat{\psi}^{\dagger}\hat{\psi}+4),
%\label{so5gammasymmradius}
\end{gather*}
and,  the radius is
\[
R_I=\frac{\alpha}{2}\sqrt{I(I+4)},
\]
where $I$ signifies an integer eigenvalue of the number operator  $\hat{I}\equiv \hat{\psi}^{\dagger}\hat{\psi}=\hat{\psi}^{\dagger}_1\hat{\psi}_1+\hat{\psi}_2^{\dagger}\hat{\psi}_2+\hat{\psi}_3^{\dagger}\hat{\psi}_3+\hat{\psi}^{\dagger}_4\hat{\psi}_4$.
Similarly, the $SO(5)$ Casimir operator is calculated as
\begin{gather*}
\sum_{a<b}\hat{X}_{ab}^2=\frac{\alpha^2}{8} (\hat{\psi}^{\dagger}\hat{\psi})(\hat{\psi}^{\dagger}\hat{\psi}+4),
%\label{so5casimirsymm}
\end{gather*}
which yields  the eigenvalue
$\frac{\alpha^2}{8}I(I+4)$, and the corresponding fully symmetric representation is constructed as
\[
|m_1,m_2,m_3,m_4\rangle=
\frac{1}{\sqrt{m_1!m_2!m_3!m_4!}}
({\hat{\psi}_1}^{\dagger})^{m_1}
({\hat{\psi}_2}^{\dagger})^{m_2}
({\hat{\psi}_3}^{\dagger})^{m_3}
({\hat{\psi}_4}^{\dagger})^{m_4}
|0\rangle ,
\]
with $m_1+m_2+m_3+m_4=I$ ($m_1,m_2,m_3,m_4\ge 0$). The dimension for the symmetric representation reads as $\frac{(I+3)!}{I! 3!}$, which is equal to that of  fuzzy $\mathbb{C}P^3$,
suggesting equivalence between fuzzy four-sphere and fuzzy $\mathbb{C}P^3$.

\section{3rd Hopf map and fuzzy manifolds}\label{sec3rdhopf}

Here, we consider  realization of the 3rd Hopf map
\begin{gather*}
S^{15} \overset{S^7}\longrightarrow S^8,
%\label{3rdcopacthopf}
\end{gather*}
 and corresponding fuzzy manifolds.   Unlike the previous cases, there are two kinds of fuzzy manifolds;
$S_F^8\simeq  SO(9)/U(4)$ and $\mathbb{C}P^7_F\simeq SU(8)/U(7)$,
depending on the choice of irreducible representation of $SO(9)$. The contents in this section are mainly based on Bernevig et al.~\cite{Bernevig2003}.

\subsection[3rd Hopf map and $SO(8)$ monopole]{3rd Hopf map and $\boldsymbol{SO(8)}$ monopole}

The 1st and 2nd Hopf maps were realized by sandwiching Pauli and quaternionic Pauli matrices by spinors. One may expect that such realization can be applied to the 3rd Hopf map. However, it is not so straightforward, since octonions cannot be represented by matrices due to their non-associative property.
To begin with,
 we  construct Majorana representation of the $SO(9)$ gamma matrix with the octonion structure constants (Table~\ref{Octoniontable}).
With $e_0=1$, the octonion algebra (\ref{splitoctonionalgebra}) is expressed as
\[
e_I e_J=-\delta_{IJ} e_0+f_{IJK}e_K,
\qquad \mbox{or}\qquad
e_P e_Q=f_{PQR}e_R,
\]
where  $P,Q,R=0,1, \dots,7$.
With use of $f_{PQR}$, the $SO(7)$ gamma matrices $-i\lambda_I$  $(I=1,2, \dots,7)$ are constructed as
\[
(\lambda_I)_{PQ}=-f_{IPQ},
\]
or
%%%%%%%%%%%%%%%%%%%%%%%%%%%%%%%%%%%%%%%%%%%%%%
\begin{alignat*}{3}
&\lambda_{1}= -i\left(
 \begin{array}{@{\,}cccc@{\,}}
 \sigma_2    &   0       &     0         &   0
\\   0        & \sigma_2 &     0         &   0
\\   0        &   0       &   \sigma_2    &   0
\\   0        &   0       &     0         & -\sigma_2
 \end{array}\right),\qquad &&
\lambda_{2} = \left(
 \begin{array}{@{\,}cccc@{\,}}
        0     & -\sigma_3 &      0        & 0
\\  \sigma_3   &   0       &      0        & 0
\\      0     &   0       &      0        & -1_2
\\      0     &   0       &  1_2    & 0
 \end{array}
 \right),& \nonumber\\
&\lambda_{3}= \left(
 \begin{array}{@{\,}cccc@{\,}}
          0   & -\sigma_1 &    0        &  0
\\  \sigma_1 &    0      &    0        &  0
\\        0   &    0      &    0        & -i\sigma_2
\\        0   &    0      & i\sigma_2   &  0
 \end{array}\right),\qquad && \lambda_{4} = \left(
 \begin{array}{@{\,}cccc@{\,}}
         0    &     0      &    -\sigma_3    &   0
\\       0    &     0      &    0      &    1_2
\\     \sigma_3    &     0      &    0      &   0
\\        0   &    -1_2     &    0      &   0
 \end{array}
 \right),& \nonumber\\
&\lambda_{5}= \left(
 \begin{array}{@{\,}cccc@{\,}}
        0    &      0      &    -\sigma_1   &    0
\\      0    &      0      &      0        &  i\sigma_2
\\ \sigma_1 &      0      &      0        &    0
\\      0    &    -i\sigma_2 &      0        &    0
 \end{array}\right) , \qquad && \lambda_{6} = \left(
 \begin{array}{@{\,}cccc@{\,}}
        0     &      0      &    0        &   -1_2
\\      0     &      0      &   -\sigma_3      &    0
\\      0     &    \sigma_3     &    0        &    0
\\     1_2    &      0      &    0        &    0
 \end{array}
 \right), & \nonumber\\
&\lambda_{7}= \left(
 \begin{array}{@{\,}cccc@{\,}}
       0      &      0      &     0       &  -i\sigma_2
\\     0      &      0      &  -\sigma_1   &    0
\\     0      & \sigma_1   &     0       &    0
\\  -i\sigma_2 &      0      &     0       &    0
\end{array} \right). &&&
\end{alignat*}
They are real antisymmetric matrices that satisfy
\[
\{\lambda_I,\lambda_J\}=-2\delta_{IJ}.
\]
With $\lambda_0\equiv 1_8$, $\lambda_0$ and $\lambda_I$ $(I=1,2, \dots,7)$ are regarded as the $SO(8)$ ``Weyl $+$'' gamma matrices.
Utilizing $\lambda_0$ and $\lambda_I$, the $SO(9)$ gamma matrices $\Gamma_A$  are constructed as
\[
\Gamma_I=i\lambda_{I}\otimes \sigma_2,\qquad \Gamma_8=1_8\otimes \sigma_1,\qquad \Gamma_9=1_8\otimes \sigma_3,
\]
or
\[
\Gamma_I=\begin{pmatrix}
0 & \lambda_{I} \\
-\lambda_{I} & 0
\end{pmatrix},\qquad
\Gamma_8=
\begin{pmatrix}
0 & 1_8 \\
1_8 & 0
\end{pmatrix},\qquad
\Gamma_9=
\begin{pmatrix}
1_8 & 0 \\
0 & -1_8
\end{pmatrix}.
\]
Again, they are real symmetric matrices that satisfy
\[
\{\Gamma_A,\Gamma_B\}=2\delta_{AB},
\]
where $A,B,C=1,2,\dots,9$.
The octonion structure constants appear in
the of\/f-diagonal elements of $\Gamma_I$.
The $SO(9)$ generators are constructed as
\begin{gather}
\Sigma_{AB}=-i\frac{1}{4}[\Gamma_A,\Gamma_B],
\label{IIso54generat}
\end{gather}
or more explicitly
%%%%%%%%%%%%%%%%%%%%%%%%%%%%%
\begin{alignat*}{3}
& \Sigma_{IJ}=\begin{pmatrix}
\sigma_{IJ} & 0 \\
0 & \sigma_{IJ}
\end{pmatrix},\qquad && \Sigma_{I8}=-\frac{i}{2}\begin{pmatrix}
\lambda_{I} & 0 \\
0 & -\lambda_{I}
\end{pmatrix},& \nonumber\\
&\Sigma_{I9}= \frac{i}{2}
\begin{pmatrix}
0 & \lambda_{I}  \\
\lambda_{I} & 0
\end{pmatrix},\qquad && \Sigma_{89}=-i\frac{1}{2}
\begin{pmatrix}
0 & -1_8 \\
1_8 & 0
\end{pmatrix},&
\end{alignat*}
where $\sigma_{IJ}$ are the $SO(7)$ generators
\[
\sigma_{IJ}=i\frac{1}{4}[\lambda_{I},\lambda_{J}].
\]
Since $\Gamma_A$ are real matrices, the corresponding $SO(9)$ generators (\ref{IIso54generat}) are purely imaginary matrices; $\Sigma_{AB}^*=-\Sigma_{AB}$. Thus, the present representation is indeed the Majorana representation, in which the charge conjugation matrix is given by  unit matrix,
and the $SO(9)$ Majorana spinor
is simply represented by (16-com\-po\-nent) real spinor.
The 3rd Hopf spinor is introduced as  $SO(9)$ Majorana spinor subject to the normalization condition
\begin{gather}
\Psi^t \Psi  =1,
\label{normalization3rdhopf}
\end{gather}
and the 3rd Hopf spinor is regarded as the coordinate of $S^{15}$.
 By sandwiching $\Gamma_A$  between the 3rd Hopf  spinors,
we now realize the 3rd  Hopf map as
\begin{gather}
\Psi\rightarrow x_A=\Psi^t \Gamma_A \Psi.
\label{map3rdhopfII}
\end{gather}
$x_A$ in (\ref{map3rdhopfII}) are coordinates on $S^8$, since
\[
\sum_{A=1,2, \dots,9}x_A x_A=(\Psi^t \Psi)^2=1.
\]
An analytic form of $\Psi$, except for the south pole, is represented as
\[
\Psi= \frac{1}{\sqrt{2(1+x_9)}}
\begin{pmatrix}
(1+x_9)
\Phi \\
(x_8-\lambda_{I} x_I) \Phi
\end{pmatrix},
\]
where $\Phi$ is a $SO(7)$ real 8-component spinor subject to the constraint
\[
\Phi^t \Phi     =1,
\]
representing the $S^7$-f\/ibre.
Then, $\Phi$ has the same degrees of freedom of  the 2nd Hopf spinor~$\psi$. We may assign  $\Phi=(\text{Re}\, \psi,\text{Im}\,\psi)^t$, and the 3rd Hopf spinor is expressed as
%%%%%%%%%%%%%%%%%%%%%%%%
\[
\Psi=
\frac{1}{\sqrt{2(1+x_9)}}
\begin{pmatrix}
(1+x_9)
\begin{pmatrix}
\text{Re} \,\psi\\
\text{Im}\, \psi
\end{pmatrix}\\
(x_8-\lambda_Ix_I)
\begin{pmatrix}
\text{Re}\, \psi\\
\text{Im}\, \psi
\end{pmatrix}
\end{pmatrix}.
\]
Naively anticipated connection $A=-i\Psi^{t}  d\Psi$ vanishes due to the Majorana property of $\Psi$, however, def\/ining
\begin{gather}
\bold{\Psi}=
\frac{1}{\sqrt{2(1+x_9)}}
\begin{pmatrix}
(1+x_9) 1_{8} \\
x_8 1_8-\lambda_{I} x_I
\end{pmatrix},
\label{nc3dhopf1}
\end{gather}
the connection of $S^7$-f\/ibre is evaluated as
\[
A=-i\bold{\Psi}^t  d\bold{\Psi}=dx_A  A_A,
\]
where $A_A=(A_M,A_9)$ ($M=1,2, \cdots,8$) are
%%%%%%%%%%%%%%%%%%%%%%%
\begin{gather}
A_{M}=-\frac{1}{1+x_9}\sigma_{MN}x_{N},\qquad
A_9=0,
\label{s08gaugefield1}
\end{gather}
with $M=1,2, \dots,8$.
Here, $\sigma_{MN}$ are $SO(8)$ ``Weyl +'' generators  given by
\[
\sigma_{IJ}=i\frac{1}{4}[\lambda_I,\lambda_J],\qquad \sigma_{I 8}=-\sigma_{8 I}=-i\frac{1}{2}\lambda_I.
\]
($\sigma_{IJ}$ and $\sigma_{I 8}$ are pure imaginary antisymmetric matrices.)
The f\/ield strength $F_{AB}=\partial_A A_B-\partial_B A_A +i[A_A,A_B]$ is also evaluated as
\begin{gather*}
F_{MN}=-x_MA_N+x_NA_M+\sigma_{MN},\qquad
F_{M9}=-F_{9M}=(1+x_9)A_M,
\end{gather*}
which represent the $SO(8)$ monopole gauge f\/ield  \cite{Grossman1984}.
Similarly, except for the north pole, the 3rd Hopf spinor has an analytic form
%%%%%%%%%%%%%%%%%%%%%%%%%%%
\begin{gather}
\bold{\Psi}'
=\frac{1}{\sqrt{2(1-x_9)}}\begin{pmatrix}
x_81_8+\lambda_I x_I\\
(1-x_9)1_8
\end{pmatrix},
\label{another3rdhopfmatrix}
\end{gather}
and the connection is
\begin{gather}
A'_{M}=-\frac{1}{1-x_9}\bar{\sigma}_{MN}x_{N},\qquad
A_9=0,
\label{s08gaugefields2}
\end{gather}
where
\[
\bar{\sigma}_{IJ}=\sigma_{IJ},\qquad \bar{\sigma}_{I8}=-\bar{\sigma}_{8I}=\sigma_{I8}.
\]
The corresponding f\/ield strength is derived as
\[
F'_{MN}=x_MA'_N-x_NA'_M+\bar{\sigma}_{MN},\qquad
F'_{M9}=-F'_{9M}=(1-x_9)A'_{M}.
\]
Here, the $SO(8)$ generators $\sigma_{MN}$  ($\bar{\sigma}_{MN}$) satisfy a  generalized self (anti-self) dual relation:
\[
\sigma_{M N}=\frac{4}{6!}\epsilon_{MN PQ ABCD}\sigma_{PQ}\sigma_{AB}\sigma_{CD},\qquad
\bar{\sigma}_{M N}=-\frac{4}{6!}\epsilon_{MN PQ ABCD}\bar{\sigma}_{PQ}\bar{\sigma}_{AB }\bar{\sigma}_{CD}.
\]
The two expressions (\ref{nc3dhopf1}) and (\ref{another3rdhopfmatrix}) are related by
\[
\bold{\Psi}'=
\begin{pmatrix}
g & 0 \\
0 & g
\end{pmatrix} \cdot \bold{\Psi},
\]
where  $g$ signif\/ies an $SO(8)$ group element
\[
g= \frac{1}{\sqrt{1-x_9^2}}(x_8+\lambda_I x_I),
\]
which yields
\[
-ig^{t}dg=-\frac{2}{1-{x_9}^2}\Sigma_{MN}^{-}x_Ndx_M,\qquad
-igdg^t=\frac{2}{1-{x_9}^2}\Sigma_{MN}^{+}x_Ndx_M.
\]
Then, the gauge f\/ields, (\ref{s08gaugefield1}) and (\ref{s08gaugefields2}),  are concisely represented as
\[
A=i\frac{1}{2}(1-x_9)dgg^t,\qquad
A'=-i\frac{1}{2}(1+x_9)g^tdg,
\]
and are related by
\[
A'=g^{t}Ag-ig^tdg.
\]
Similarly, their f\/ield strengths are
\[
F'=g^tFg.
\]
The non-trivial topological structure of the $SO(8)$ monopole bundle is guaranteed by the homotopy theorem
\[
\pi_{7}(SO(8))\simeq \mathbb{Z}\oplus \mathbb{Z},
\]
which is specif\/ied by
the Euler number
\[
e=
\int_{S^8} {\rm tr}\,(F\wedge F\wedge F\wedge F).
\]

\subsection[Fuzzy $\mathbb{C}P^7$ and fuzzy $S^8$]{Fuzzy $\boldsymbol{\mathbb{C}P^7}$ and fuzzy $\boldsymbol{S^8}$}

In the realization of the 3rd Hopf map, we utilized the real (Majorana) spinor.
The fuzzif\/ication  procedure in the previous sections can {\it not} be  straightforwardly applied to the present case, since we do not have the  complex conjugate spinor to be identif\/ied with derivative.
However, with 16 real components of the 3rd Hopf spinor, we can construct an 8-component normalized complex spinor to be identif\/ied with
 coordinates on~$\mathbb{C}P^7$.  Then, in the present case,
 there exist two dif\/ferent types of fuzzy manifolds, depending on the choice of the irreducible representation of~$SO(9)$. The f\/irst one is the above mentioned fuzzy  $\mathbb{C}P^7\simeq S^{15}/S^1$ specif\/ied by the vector representation of $SO(9)$, while the other is the fuzzy eight-sphere $S_F^8\simeq SO(9)/U(4)$ specif\/ied by the spinor representation.
Both two  fuzzy manifolds are reasonable generalizations of the previous low dimensional fuzzy spheres and fuzzy complex projective spaces
(see Table~\ref{Tablehopfmapfuzzysphere}).

%%%%%%%%%%%%%%%%%%%%%
\begin{table}
\centering
 \caption{In the 1st and 2nd Hopf maps, the fuzzy manifolds are uniquely determined, since  the corresponding fuzzy spheres and complex projective spaces are equivalent, i.e.\ $S_F^2\simeq \mathbb{C}P^1_F$ and $S_F^4\simeq \mathbb{C}P^3_F$. Meanwhile, in the 3rd Hopf map, two corresponding fuzzy manifolds, $S_F^8\simeq SO(9)/U(4)$ and $\mathbb{C}P^7_F\simeq S^{15}/S^1$, are not equivalent.}
 \label{Tablehopfmapfuzzysphere}

 \vspace{1mm}

\begin{tabular}{|c||c|c|c|}
\hline   Division algebra & Complex numbers  & Quaternions & Octonions
\\
\hline Hopf maps  &  1st
& 2nd  & 3rd  \\
\hline Fuzzy sphere  & $S_F^2\simeq SO(3)/U(1)$& $S_F^4\simeq SO(5)/U(2)$ &  $S_F^8\simeq SO(9)/U(4)$ \\
\hline
Fuzzy $\mathbb{C}P^n$ & $\mathbb{C}P^1_F%\simeq SU(2)/U(1)
\simeq S^3/S^1$   &  $\mathbb{C}P^3_F%\simeq SU(2)/U(1)
\simeq S^7/S^1$   & $\mathbb{C}P^7_F%\simeq SU(2)/U(1)
\simeq S^{15}/S^1$
\\
\hline
\end{tabular}
\end{table}

\section{Beyond Hopf maps: even higher dimensional generalization}\label{sechbeyondhopf}

We have reviewed the construction of  fuzzy manifolds based on the Hopf maps.
Since the Hopf maps are only three kinds, the corresponding fuzzy manifolds are also limited. However, from the results of the Hopf maps,  one may naturally infer two possible generalizations of the fuzzy manifolds, one of which is a series of fuzzy spheres:
\[
S_F^{2k}\simeq SO(2k+1)/U(k),
\]
 and the other is that of  fuzzy complex projective spaces:
\[
 \mathbb{C}P^k_F\simeq SU(k+1)/U(k)\simeq S^{2k+1}/S^1.
\]
In this section, we f\/irst introduce  mathematics of fuzzy spheres $S^{2k}_F$ in arbitrary even dimensions, and next provide their physical interpretations, mainly based on Hasebe and Kimura~\cite{hasebekimurahep-th/0310274} (see also Fabinger~\cite{Fabinger2002} and Meng~\cite{GuowuMeng2003}).
Fuzzy $\mathbb{C}P^k$ manifolds and their corresponding Landau models  are discussed in Appendix~\ref{sectioncomplexprojective}.

\subsection{Clif\/ford algebra: another generalization of complex numbers}

It may be worthwhile to begin with the story of  generalization of the complex numbers to Clif\/ford algebra. Generalization from fuzzy two-sphere to its higher dimensional cousins are quite analogous to the generalization  of complex numbers.
As discussed in Section~\ref{hopfmapdivisionalgebra},
 Cayley--Dickson construction provides one systematic way to construct new numbers by duplicating the original numbers, but this construction method  has a fatal problem: If we utilize the method, the resulting algebra loses  a nice property of numbers one by one.
 For instance, in the construction of quaternions, the commutativity of the complex numbers was lost. In the construction of octonions, even the associativity of  quaternions was abandoned.
 Consequently, generalization of complex numbers ends up with the octonions, and there are only three division algebras (except for real numbers).
Clif\/ford found another generalization of complex numbers, based on  Hamilton's quaternions and Grassmann algebras, known as Clif\/ford algebra.
The Clif\/ford algebra,
  $\text{Clif\/f}_{n}$, consists of $2^n$ basis elements $\{1,e_a,e_a e_b,e_a e_b e_b, \dots,e_1 e_2 e_3 \cdots e_n\}$  $(a\neq b, a\neq b\neq c, \dots)$, and its algebraic structure is determined by the relation\footnote{Though we treat Clif\/ford algebras with Euclidean signature,  Clif\/ford algebras can be generally def\/ined with indef\/inite signature.}
\begin{gather}
\{e_a,e_b\}=2\delta_{ab}.
\label{Cliffordalgebra}
\end{gather}
The division algebras except for the octonions are realized as special cases of Clif\/ford algebra, i.e.\  $\mathbb{R}=\text{Clif\/f}_{0}$, $\mathbb{C}=\text{Clif\/f}_{1}$, and $\mathbb{H}=\text{Clif\/f}_{2}$.   The Clif\/ford algebra  can also be regarded as the ``quantized'' Grassmann algebra (compare (\ref{Cliffordalgebra}) with $\{\eta_{a},\eta_b\}=0$ (\ref{algebraofgrassmann})).
 Though the division property is in general lost, the Clif\/ford algebras  always maintain the nice associative property, and are represented by gamma matrices that satisfy
\[
\{\gamma_a,\gamma_b\}=2\delta_{ab}.
\]
Importantly, there are analogous geometrical properties between the division algebras and the Clif\/ford algebras: In the division algebras, new numbers are constructed by the Cayley--Dickson construction.
 Similarly, higher dimensional gamma matrices are constructed by the lower dimensional gamma mat\-ri\-ces.
Specif\/ically, $SO(2k-1)$ gamma mat\-ri\-ces $\gamma_i^{(2k-1)}$ $(i=1,2, \dots,2k-1)$ are provided, $SO(2k+1)$ gamma matrices
$\gamma_a^{(2k+1)}$ $(a=1,2, \dots,2k+1)$ can be constructed as
%%%%%%%%%%%%%%%%%%%%%%%%
\begin{gather}
\gamma_i^{(2k+1)}=
\begin{pmatrix}
0 & i\gamma_i^{(2k-1)} \\
-i\gamma_i^{(2k-1)} & 0
\end{pmatrix},\qquad \gamma_{2k}^{(2k+1)}=
\begin{pmatrix}
0 &  1 \\
1 & 0
\end{pmatrix},\qquad \gamma_{2k+1}^{(2k+1)}=
\begin{pmatrix}
1 &  0 \\
0 & -1
\end{pmatrix}.
\label{constructionofgamma}
\end{gather}
Thus,  in any higher dimensional  gamma matrices are constructed by repeating the above procedure from the $SO(3)$ gamma matrices, $\gamma_1^{(3)}=-\sigma_2$, $\gamma_2^{(3)}=\sigma_1$, $\gamma_3^{(3)}=\sigma_3$.
As the Hopf maps exhibit the hierarchical structure stemming from the Cayley--Dickson construction,
the geometry of higher dimensional fuzzy spheres ref\/lects the iterative construction structure of the gamma matrices as we shall see below.

\subsection{Mathematical aspects of fuzzy sphere}

We f\/irst introduce mathematical construction of   fuzzy spheres \cite{hep-th/9602115,hep-th/9712105,Ho2002,Kimura2002,hep-th/0105006,AzumaBagnoud2003,Abe2004}.
Coordinates on fuzzy $2k$-sphere are constructed by the $SO(2k+1)$ gamma matrices in the fully symmetric representation $[\frac{I}{2},\frac{I}{2},\dots,\frac{I}{2}]$:
\begin{gather*}
X_a=\frac{\alpha}{2}(\gamma_a\otimes 1 \otimes  \cdots \otimes 1+\gamma_a\otimes 1 \otimes  \cdots \otimes 1+ \cdots+1 \otimes  \cdots \otimes 1\otimes \gamma_a)_{\rm sym},
%\label{symmetricprodX}
\end{gather*}
%%%%%%%%%%%%%%%%%%%%%%%%%%%%%%%%%%%%%
where $\gamma_a$ $(a=1,2,\dots,2k+1)$ are the $SO(2k+1)$ gamma matrices in the fundamental representation, and the number of the tensor product is $I$.
Square of the radius of fuzzy $2k$-sphere is given by
\[
\sum_{a=1}^{2k+1}X_aX_a=\frac{\alpha^2}{4}I(I+2k).
\]
Similarly, in the symmetric representation, the eigenvalue of the $SO(2k+1)$ Casimir is
\[
\sum_{a<b}^{2k+1}X_{ab}^2=\frac{\alpha^2}{16}{k}I(I+2k),
\]
where $X_{ab}=-i\frac{1}{4}[X_a,X_b]$ are the $SO(2k+1)$ generators.
(Detail calculation techniques for symmetric representation can be found in~\cite{Azumathesis}.)
Thus, the index $I$ of the symmetric representation determines the magnitude of the radius of fuzzy sphere.
As the dimension of the symmetric representation becomes ``larger'', the corresponding fuzzy sphere becomes ``larger''.

As in the case of fuzzy-four sphere, the fuzzy coordinates $X_a$  do not satisfy a closed algebra by themselves, but $X_a$ and $X_{ab}$ satisfy the following enlarged algebra
\begin{gather*}
[X_a,X_b]=2i\alpha X_{ab},\qquad
[X_a,X_{bc}]= -i\frac{\alpha}{2}(\delta_{ab}X_c-\delta_{ac}X_b),\nonumber\\
[X_{ab},X_{cd}]=i\frac{\alpha}{2}(\delta_{ac}X_{bd}-\delta_{ad}X_{bc}+\delta_{bc}X_{ad}-\delta_{bd}X_{ac}).
\end{gather*}
With identif\/ication $X_{a, 2k+2}=-X_{2k+2,a}=\frac{1}{2}X_a$ and  $X_{ab}=X_{ab}$,
the above algebra is found to be equivalent to the $SO(2k+2)$ algebra
\[
[X_{AB},X_{CD}]=i\frac{\alpha}{2}(\delta_{AC}X_{BD}-\delta_{AD}X_{BC}+\delta_{BC}X_{AD}-\delta_{BD}X_{AC}),
\]
where $A,B,C,D=1,2,\dots,2k+2$. Thus, the algebra of  fuzzy sphere $S^{2k+1}_F$ is $SO(2k+2)$~\cite{Ho2002}, and
$S^{2k}_F$ is expressed as
\[
S_F^{2k}\simeq SO(2k+2)/U(k+1)\simeq SO(2k+1)/U(k),
\]
where we used the relation
$SO(2k+2)/SO(2k+1)\simeq S^{2k+1}\simeq U(k+1)/U(k)$.
 The above coset representation can also be expressed as
\begin{gather}
S_F^{2k}\approx SO(2k+1)/SO(2k)\times SO(2k-1)/U(k-1)\approx S^{2k}\times S_F^{2k-2}.
\label{locallyhierarchy2ksphere1}
\end{gather}
Fuzzy $2k$-sphere is  constructed not only by the operators $X_a$ but also
 the ``extra'' operators $X_{ab}$, and the very existence of $X_{ab}$ brings the extra fuzzy-space $S_F^{2k-2}$ over $S^{2k}$.
%As the $SO(2k+1)$ gamma matrix is constructed by the $SO(2k-1)$ gamma matrix (\ref{constructionofgamma}), the geometry of the fuzzy $2k$-sphere %ref\/lects the dimensional structure of  gamma matrices.
From (\ref{locallyhierarchy2ksphere1}), one may f\/ind the fuzzy sphere is expressed by the hierarchical f\/ibrations of lower dimensional spheres
\begin{gather}
S_{F}^{2k}\approx S^{2k}\times S^{2k-2}\times  \cdots \times S^4\times S^2,
\label{locallyhierarchy2ksphere2}
\end{gather}
%%%%%%%%%%%%%%%%%%%%%%
which ref\/lects the iterative construction of gamma matrices from lower dimensions.

\subsection[Hopf spinor matrix and $SO(2k)$ monopole]{Hopf spinor matrix and $\boldsymbol{SO(2k)}$ monopole}

To obtain monopole bundles in generic even dimensions,
we ``extend'' the Hopf maps.
First, we def\/ine  ``Hopf spinor matrix'' of the form
\begin{gather}
\bold{\Psi}=\frac{1}{\sqrt{2(1+x_{2k+1}})}
\begin{pmatrix}
(1+x_{2k+1}) \bold{1} \\
(x_{2k}\bold{1}  - i\gamma_i x_i)
\end{pmatrix},
\label{generalizedHopfspinor}
\end{gather}
where  $\bold{1}$ stands for
 $2^k\times 2^k$ unit matrix, $\gamma_i$ $(i=1,2, \dots, 2k-1)$ are   $SO(2k-1)$ gamma matrices, and $x_a$ $(a=1,2,\dots,2k+1)$ are coordinate on $2k$-sphere satisfying $\sum_{a=1}^{2k+1}x_a x_a =1$.
The Hopf spinor matrix is a $2^{k+1}\times 2^{k}$ matrix that satisf\/ies
\[
x_{a} \bold{1} =\bold{\Psi}^{\dagger}\gamma_a \bold{\Psi},
\]
where $\gamma_a$ are $SO(2k+1)$ gamma matrices (\ref{constructionofgamma}).
The corresponding monopole gauge f\/ield is evaluated as
\[
A=dx_a A_a =-i\bold{\Psi}^{\dagger}d\bold{\Psi},
\]
where
\begin{gather}
A_{\mu}=-\frac{1}{2(1+x_{2k+1})} \Sigma_{\mu\nu}^+ x_{\nu},\qquad A_{2k+1}=0.
\label{so2kgauge1field}
\end{gather}
$\Sigma_{\mu\nu}^{+}$ ($\mu,\nu=1,2, \dots,2k$) are  $SO(2k)$ generators given by
\[
\Sigma_{ij}^+=-i\frac{1}{4}[\gamma_i,\gamma_j],\qquad \Sigma_{i 2k}^+=-\Sigma_{2k i}^+=\frac{1}{2}\gamma_i.
\]
The f\/ield strength $F_{ab}=\partial_aA_b-\partial_bA_a+i[A_a,A_b]$ is calculated  to yield
\[
F_{\mu\nu}=-x_{\mu}A_{\nu}+x_{\nu}A_{\mu}+\Sigma_{\mu\nu}^{+},\qquad
F_{\mu, 2k+1}=-F_{2k+1, \mu}=(1+x_9)A_{\mu}.
\]
These are the $SO(2k)$ non-Abelian monopole gauge f\/ield strength~\cite{Horvath1978,Tchrakian1980,Saclioglu1986}.
Another representation of the Hopf spinor matrix is introduced as
\begin{gather}
\bold{\Psi}'=\frac{1}{\sqrt{2(1-x_{2k+1}})}
\begin{pmatrix}
(x_{2k}\bold{1}+i\gamma_i x_i) \\
(1-x_{2k+1})\bold{1}
\end{pmatrix},
\label{hopfspinormatrix2}
\end{gather}
and the corresponding gauge f\/ield, $A'=-i\bold{\Psi}^{\dagger}d\bold{\Psi}'$, reads as
\begin{gather}
A'_{\mu}=-\frac{1}{2(1-x_{2k+1})} \Sigma_{\mu\nu}^{-}x_{\nu},\qquad A'_{2k+1}=0,
\label{so2kgauge2field}
\end{gather}
where
\[
\Sigma_{ij}^{-}=\Sigma_{ij}^{+},\qquad \Sigma^{-}_{i,2k}=-\Sigma_{2k,i}^{+}.
\]
The gauge f\/ield strength $F'_{ab}$ is
\[
F'_{\mu\nu}=x_{\mu}A'_{\nu}-x_{\nu}A'_{\mu}+\Sigma_{\mu\nu}^{-},\qquad
F'_{\mu, 2k+1}=-F'_{2k+1, \mu}=(1-x_9)A'_{\mu}.
\]
The $SO(2k)$ generators, $\Sigma_{\mu\nu}^+$ and $\Sigma_{\mu\nu}^-$, satisfy the generalized self and anti-self dual relations, respectively:
\[
\Sigma_{\mu_1\mu_2}^{\pm}=\pm\frac{2^{k-1}}{(2k-2)!}\epsilon_{\mu_1\mu_2\mu_3\mu_4 \cdots\mu_{2k-1}\mu_{2k}}\Sigma_{\mu_3\mu_4}^{\pm} \cdots\Sigma_{\mu_{2k-1}\mu_{2k}}^{\pm}.
\]
The two Hopf spinor matrices (\ref{generalizedHopfspinor}) and (\ref{hopfspinormatrix2}) are related by
the transformation
\[
\bold{\Psi}'=
\begin{pmatrix}
 g & 0 \\
 0 & g
\end{pmatrix}
 \cdot \bold{\Psi},
\]
where $g$  is an $SO(2k)$ group element
\[
g=\frac{1}{\sqrt{1-x_{2k+1}^2}}(x_{2k}+i\gamma_ix_i).
\]
With  $g$, the gauge f\/ields (\ref{so2kgauge1field}) and (\ref{so2kgauge2field}) are concisely represented as
\[
A=i\frac{1}{2}(1-x_{2k+1})dg g^{\dagger},\qquad
A'=-i\frac{1}{2}(1+x_{2k+1})g^{\dagger}dg,
\]
where
\[
-ig^{\dagger}dg=-\frac{2}{1-x_{2k+1}^2}\Sigma_{\mu\nu}^{-}x_{\nu}dx_{\mu},\qquad
-igdg^{\dagger}=\frac{2}{1-x_{2k+1}^2}\Sigma_{\mu\nu}^{+}x_{\nu}dx_{\mu}.
\]
Then, the two gauge f\/ields are  related as
\[
A'=g^{\dagger}Ag-ig^{\dagger}dg,
\]
and the f\/ield strengths are
\[
F'=g^{\dagger}F g.
\]
The homotopy theorem guarantees the non-trivial topology of the $SO(2k)$ bundle fibration over~$S^{2k}$:
\[
\pi_{2k-1}(SO(2k))=\mathbb{Z},
\]
which is specif\/ied by the Euler number
\[
e=\int_{S^{2k}} {\rm tr}\,\big(F^k\big).
\]

\subsection[$SO(2k+1)$ Landau model]{$\boldsymbol{SO(2k+1)}$ Landau model}

In  generic $d$-dimensional space, Landau Hamiltonian is given by~\cite{hasebekimurahep-th/0310274}
\begin{gather}
H=-\frac{1}{2M}D_a^2=-\frac{1}{2M}\frac{\partial^2}{{\partial r}^2}-(d-1)\frac{1}{2Mr}\frac{\partial}{\partial r}+\frac{1}{2Mr^2}\sum_{a<b}\Lambda_{ab}^2,
\label{LandauHamilgeneric}
\end{gather}
where
$D_a=\partial_a+iA_a$ ($A_a$ is the $SO(2k)$ monopole gauge f\/ield), and
$\Lambda_{ab}=-ix_aD_b+ix_bD_a$.
The covariant momentum $\Lambda_{ab}$ does not satisfy the $SO(2k+1)$ algebra but satisf\/ies
\begin{gather*}
[\Lambda_{ab},\Lambda_{cd}]= i(\delta_{ac}\Lambda_{bd}+\delta_{bd}\Lambda_{ac}-\delta_{bc}\Lambda_{ad}-\delta_{ad}\Lambda_{bc})\\
\hphantom{[\Lambda_{ab},\Lambda_{cd}]=}{}
-i(x_a x_c F_{bd}+x_{b}x_{d}F_{ac}-x_{b}x_{c}F_{ad}-x_{a}x_{d}F_{bc} ).
\end{gather*}
The conserved $SO(2k+1)$ angular momentum is constructed as
\[
L_{ab}=\Lambda_{ab}+r^2F_{ab},
\]
which satisf\/ies the $SO(2k+1)$ algebra
\[
{}[L_{ab},L_{cd}]= i(\delta_{ac}L_{bd}+\delta_{bd}L_{ac}-\delta_{bc}L_{ad}-\delta_{ad}L_{bc}),
\]
and generates the $SO(2k+1)$ transformations, for instance
\begin{gather*}
[L_{ab},\Lambda_{cd}]= i(\delta_{ac}\Lambda_{bd}+\delta_{bd}\Lambda_{ac}-\delta_{bc}\Lambda_{ad}-\delta_{ad}\Lambda_{bc}),\nonumber\\
[L_{ab},F_{cd}]= i(\delta_{ac}F_{bd}+\delta_{bd}F_{ac}-\delta_{bc}F_{ad}-\delta_{ad}F_{bc}).
\end{gather*}
On  $2k$-sphere, the Landau Hamiltonian (\ref{LandauHamilgeneric}) is reduced to  $SO(2k+1)$ Landau Hamiltonian
\[
H=\frac{1}{2MR^2} \sum_{a<b}\Lambda_{ab}^2.
\]
With the orthogonality $\Lambda_{ab}F_{ab}=F_{ab}\Lambda_{ab}$, the Hamiltonian is rewritten as
\[
H=\frac{1}{2MR^2}\left(\sum_{a<b}L_{ab}^2-\sum_{a<b}F_{ab}^2\right)=\frac{1}{2MR^2}\left(\sum_{a<b} L_{ab}^2-\sum_{\mu<\nu}\Sigma_{\mu\nu}^2\right),
\]
where $\sum_{a<b}F_{ab}^2=\sum_{\mu<\nu}\Sigma_{\mu\nu}^2$ was used.
As in the previous Landau models,  we take  the fully symmetric spinor representation $({I}/{2})\equiv [\frac{I}{2},\frac{I}{2}, \dots,\frac{I}{2}]$ for the $SO(2k)$ Casimir $\sum_{\mu<\nu} \Sigma_{\mu\nu}^2$, and
 the irreducible representation   $(n,I/2)\equiv[n+\frac{I}{2},\frac{I}{2},\frac{I}{2}, \dots,\frac{I}{2}]$ for the $SO(2k+1)$ Casimir $\sum_{a<b}L_{ab}^2$, with the Landau level index~$n$.
In such a representation, the energy eigenvalues are given by
\[
E_n=\frac{1}{2MR^2}(C_{2k+1}(n,{I}/{2})-C_{2k}({I}/{2})),
\]
where $C_{2k+1}(n,{I}/{2})$ is
the $SO(2k+1)$ Casimir eigenvalue for $(n,I/2)$:
\[
C_{2k+1}(n,{I}/{2})=n^2+n(I+2k-1)+\frac{k}{4}I(I+2k),
\]
and $C_{2k}({I}/{2})$ is the $SO(2k)$ Casimir eigenvalue for $({I}/{2})$:
\[
C_{2k}({I}/{2})=\sum_{\mu < \nu} \Sigma_{\mu\nu}^2 =\frac{1}{4}Ik(I+2k-2).
\]
Consequently, the energy eigenvalues of the $SO(2k+1)$ Landau model are  derived as
\[
E_n=\frac{1}{2MR^2} (n^2+n(I+2k-1)+\frac{1}{2}Ik).
\]
In the thermodynamic limit: $R,I\rightarrow \infty$ with $I/R^2$ f\/ixed, the Landau levels are reduced to
\[
E_n\rightarrow \frac{I}{2MR^2}(n+\frac{1}{2}k).
\]
The LLL energy, $E_{\rm LLL}=\frac{I}{4MR^2}k$, depends on the spacial dimension $2k$.
 In the thermodynamic limit, $S^{2k}$ is reduced to $2k$-dimensional plane, and the zero-point energy $\frac{B}{2M}=\frac{I}{4MR^2}$ coming from each 2-dimensional plane amounts to $E_{\rm LLL}$.

As discussed above,  coordinates on fuzzy $2k$-sphere are given by the $SO(2k+1)$ gamma matrices in the symmetric spinor representation.
Similarly, the LLL basis of the $SO(2k+1)$ Landau model realizes such a symmetric spinor representation.
Thus, the LLL of $SO(2k+1)$ Landau model provides a  physical set-up for $2k$-dimensional fuzzy sphere
(see Table~\ref{correspondencefuzzyandmonopole}).

\begin{table}[t]
\centering
 \caption{The fuzzy $2k$-sphere is physically realized in the LLL of the   $SO(2k+1)$ Landau model.
  Previously encountered monopoles  are understood as the special cases of $SO(2k)$ monopoles, for instance $U(1)\simeq SO(2)$, $SU(2)(\otimes SU(2))\simeq SO(4)$. Note also $SU(4)\simeq SO(6)$.  The holonomy groups of spheres are equal to the corresponding monopole gauge groups.}
 \label{correspondencefuzzyandmonopole}

 \vspace{1mm}

\begin{tabular}{|c|c|c|}
\hline   Fuzzy sphere & Corresponding original sphere  &  Monopole gauge group
\\
\hline\hline $S_F^2\simeq SO(3)/U(1)$  &  $S^2\simeq SO(3)/SO(2)$
& $U(1)$  \\
\hline $S_F^4\simeq SO(5)/U(2)$  & $S^4\simeq SO(5)/SO(4)$ & $SU(2)$  \\
\hline
$S_F^6\simeq SO(7)/U(3)$ & $S^{6}\simeq SO(7)/SO(6)$  &   $SU(4)$
\\
\hline
$S_F^8\simeq SO(9)/U(4)$ & $S^{8}\simeq SO(9)/SO(8)$  &   $SO(8)$
\\
 \hline
 \vdots &  \vdots &  \vdots
\\
\hline
$S_F^{2k}\simeq SO(2k+1)/U(k)$ & $S^{2k}\simeq SO(2k+1)/SO(2k)$ & $SO(2k)$  \\
\hline
\end{tabular}
\end{table}

\subsection{Dimensional hierarchy}

\begin{figure}[t]
\centerline{\includegraphics[width=58mm]{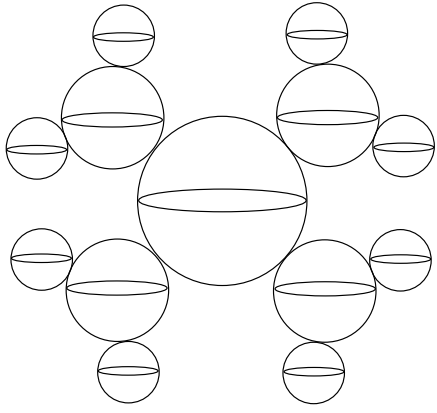}}
\caption{Lower dimensional spheres gather spherically
 to form a higher dimensional fuzzy sphere.}
\label{DimHierarchy}
\end{figure}

 Here, we give a physical interpretation of the hierarchical geometry of
 higher dimensional fuzzy spheres (\ref{locallyhierarchy2ksphere2}).
From the formula of the irreducible representation of $SO(2k+1)$~\cite{IachelloBook}, the degeneracy in  $n$th LL is given by
\begin{gather*}
d(n)=\frac{2n\!+\!I\!+\!2k\!-\!1}{(2k\!-\!1)!!}\frac{(n\!+\!k\!-\!1)!}{n!(k\!-\!1)!}\prod_{i=1}^{k-1}(I\!+\!2i\!-\!1)\prod_{i=2}^k \frac{n\!+\!I\!+\!2k\!-\!i}{2k\!-\!i}\prod_{l=1}^{k-2}\prod_{i=l+2}^k \frac{I\!+\!2k\!-\!i\!-\!l}{Ik\!-\!i\!-\!l}.
\end{gather*}
In the LLL, the representation is reduced to the fully symmetric spinor representation $[\frac{I}{2},\!\frac{I}{2},\! {\dots},\!\frac{I}{2}]$,  and the degeneracy becomes
\begin{gather}
d_{\rm LLL}=\frac{(I+2k-1)!!}{(2k-1)!!(I-1)!!}
\prod_{l=1}^{k-1}\frac{(I+2l)!l!}{(I+l)!(2l)!} \ \longrightarrow \ I \cdot I^2 \cdot I^3 \cdots I^{k-1} \cdot I^{k}=I^{\frac{1}{2}k(k+1)}.
\label{iterativedegeneracy}
\end{gather}
The last expression implies a nice intuitive picture of the hierarchical geometry of fuzzy spheres.
Each of the $SO(2k)$ monopole f\/luxes on $S^{2k}$ occupies an area $\ell_B^{2k}=(1/B)^k={(2R^2/I)}^{k}$ with the magnetic f\/ield $B=(2\pi I)/(4\pi R^2)$, and the number of f\/luxes on $S^{2k}$ is  $\sim R^{2k}/\ell_B^{2k}\sim I^k $.
Besides, the monopole f\/lux itself is represented by the generators of the non-Abelian $SO(2k)$ group, and is regarded as a $(2k-2)$-dimensional fuzzy sphere. Again, the $(2k-2)$-dimensional fuzzy sphere is interpreted as a
 $(2k-2)$-dimensional sphere in $SO(2k-2)$ monopole background, and then, on $S^{2k-2}$, there are $SO(2k-2)$ f\/luxes each of which occupies the area $\ell_B^{2k-2}=(1/B)^{k-1}={(2R^2/I)}^{k-1}$. Therefore, the number of $SO(2k-2)$ f\/luxes on $S^{2k-2}$ is $\sim R^{2k-2}/\ell_B^{2k-2}\sim I^{k-1} $.
Similarly, the $SO(2k-2)$ non-Abelian f\/lux is given by the generators of the $SO(2k-2)$ group, and regarded as a fuzzy $S^{2k-4}$.  Thus,  on $S^{2k}$, we have $I^k$ $S^{2k-2}$,  on which $I^{k-1}$ $S^{2k-4}$, on which $I^{k-2}$ $S^{2k-6}$, \dots. By this iteration, we obtain the formula~(\ref{iterativedegeneracy}).  Inversely, we can view this mechanism from low dimensions: Lower dimensional spheres gather spherically to  form  a higher dimensional sphere, and such iterative process amounts to construct a higher dimensional fuzzy sphere. The dimensional hierarchy is depicted in Fig.~\ref{DimHierarchy}.

\section{Summary and discussions}\label{secsummary}

\begin{table}[t]
\centering
 \caption{Correspondence between algebras, monopoles, and fuzzy spheres.}
 \label{algebramaomonopole}
 \vspace{1mm}

\begin{tabular}{|@{\,}c@{\,}||@{\,}c@{\,}|@{\,}c@{\,}|@{\,}c@{\,}|}
\hline
   Algebras &   Bundle structure & Monopoles  & Fuzzy manifolds \\
\hline \hline
   Division algebra &  Canonical bundle & $U(1)$, $SU(2)$, $SO(8)$  &  Fuzzy 2,4,8-spheres \\
   \hline
   Grassmann algebra &   Graded canonical bundle & Supermonopole & Fuzzy supersphere   \\
 \hline
   Clif\/ford algebra &  Spinor bundle         &  $SO(2k)$  & Higher d. fuzzy spheres \\
\hline
\end{tabular}
\end{table}

 We reviewed the close relations between monopoles, LLL, and fuzzy spheres.  The fuzzy $2k$-sphere is physically realized in the LLL of the $SO(2k+1)$ Landau model.
 In the generalization of fuzzy spheres,
 three classical algebras; division algebra, Grassmann algebra,  and Clif\/ford algebra, played crucial roles. They brought the basic structures of  monopole bundles and fuzzy spheres (Table \ref{algebramaomonopole}).
In particular, the hierarchical geometry of fuzzy spheres  is a direct manifestation of their gamma matrix construction: As higher dimensional gamma matrices are constructed by lower  dimensional gamma matrices,
higher dimensional fuzzy sphere is constructed by lower dimensional spheres. Such dimensional hierarchy can be physically understood in the context of higher dimensional Landau model.  It should  be mentioned that,
  in~\cite{hep-th/0402044},
such interpretation was successfully applied to dual description of higher dimensional fuzzy spheres in string theory.
We also emphasize the importance of the
 Hopf map in realizing fuzzy sphere.
In the fuzzif\/ication of spheres, the total manifolds (Hopf spinor spaces) of the Hopf maps played a fundamental role: The total manifolds were f\/irstly fuzzif\/icated and as ``a~consequence'' the basemanifolds (spheres) are fuzzif\/icated.
Interestingly, this fuzzif\/ication mechanism coincides with the philosophy of  twistor theory (see~\cite{arXiv:0902.2523} and references therein).

There are various works related to the present paper: Even restricted to
the recent ones,
  supersymmetric extensions of the Landau model  \cite{arXiv:hep-th/0404108,arXiv:hep-th/0503162,arXiv:hep-th/0510019,arXiv:hep-th/0612300,arXiv:0806.4716,arXiv:1003.0218}, supersymmetric fuzzy manifolds \cite{arXiv:hep-th/0311159,arXiv:hep-th/0611328},
 generalizations of  the Hopf maps \cite{math/0407342,arXiv:0905.2792,arXiv:0912.3279,arXiv:1008.2589}, supersymmetric quantum mechanics \cite{hep-th/0606152,arXiv:0902.2682,arXiv:0905.3461,arXiv:0905.4951,arXiv:0911.3257,arXiv:0912.3289,arXiv:1004.4597,arXiv:1001.2659}, and applications to string theory \cite{arXiv:hep-th/0703021,arXiv:0801.1813,arXiv:0903.3966,arXiv:0908.3263}.

 Finally, we comment on applications to many-body physics.
  The correspondence between fuzzy geometry and LLL physics argued in the paper was at one-particle level observation.
Interestingly, there even exists correspondence at many-body level:  Many-body groundstate wavefunction of quantum Hall ef\/fect (Laughlin wavefunction) is mathematically analogous to an antiferromagnetic ground state (AKLT state)~\cite{Arovas1988,arXiv:0901.1498}.
Accompanied with the higher dimensional and supersymmetric generalizations of the quantum Hall ef\/fect\footnote{Interested readers are expected to consult the review~\cite{hep-th/0606161}, and~\cite{arXiv:hep-th/0411137,arXiv:hep-th/0504092,arXiv:hep-th/0505095,arXiv:0809.4885,arXiv:0902.2523} for more recent works.}, their formalism has begun to be applied to the construction of antiferromagnetic quantum spin states with higher symmetries~\mbox{\cite{arXiv:0901.1498, Tuetal2009}}.

\appendix

%\section*{Appendix}\label{seccpspace}

\section[0th Hopf map and $SO(2)$ ``Landau model'']{0th Hopf map and $\boldsymbol{SO(2)}$ ``Landau model''}\label{sect0thhopf}

Real numbers are the ``0th'' member of the division algebra. For completeness,
 we introduce the 0th Hopf map
\[
S^1 \overset{Z_2}\longrightarrow S^1,
%\label{0thcopacthopf}
\]
and the corresponding Landau model.
The 0th Hopf map
is realized by identifying ``opposite points'' on a circle, and is simply visualized as the geometry of M\"obius strip whose basemanifold is $S^1$ and transition function  is $Z_2$.  Unlike the other Hopf maps, the dimension of the basemanifold
 is odd and the structure group is a discrete group.

\subsection{Realization of the 0th Hopf map}

With the coordinate on a circle, $w=(w_1,w_2)^t$ (a real two-component  spinor subject to $w^tw=w_1^2+w_2^2=1$),
the 0th Hopf map is realized as
\begin{gather}
w=
\begin{pmatrix}
w_1\\
w_2
\end{pmatrix}
\rightarrow x_1=w^t\sigma_3 w,\qquad x_2=w^t\sigma_1 w.
\label{expli0thhopf}
\end{gather}
%%%%%%%%%%%%%%%%%%%%
Here, $\sigma_1$ and $\sigma_3$ are the real Pauli matrices.
$x_1$ and $x_2$ are invariant under the transformation $(w_1,w_2) \rightarrow -(w_1,w_2)$, and hence $(x_1,x_2)$ is $Z_2$ projection of $(w_1,w_2)$.
From~(\ref{expli0thhopf}), one may f\/ind  $x_1^2+x_2^2=(w^tw)^2=1$.
Inverting the map,
 $w$ can be expressed as
\[
w
=
 \frac{1}{\sqrt{2(1+x_1)}}
\begin{pmatrix}
1+x_1\\
x_2
\end{pmatrix}=
\begin{pmatrix}
 \cos\dfrac{\theta}{2}\vspace{1mm}\\
\sin\dfrac{\theta}{2}
\end{pmatrix},
\]
where $x_1$ and $x_2$ are  parameterized as $x_1= \cos\theta$ and $y=\sin\theta$.
The corresponding connection vanishes: $A=-iw^t dw =0$.
Meanwhile, using a $U(1)$ element $\omega=w_1+iw_2=e^{i\frac{\theta}{2}}$, the 0th Hopf map is restated as
\[
\omega\rightarrow x_1+ix_2= \omega^2,
\]
and the connection, $A=-i\omega^*d\omega=iw^t\sigma_2dw=dx A_x +dy A_y$,  is given by
\[
A_x=-\frac{1}{2}y,\qquad A_y=\frac{1}{2}x
\qquad \mbox{or}\qquad
A_{\theta}=-\frac{1}{2}.
\]
It is straightforward to see the f\/ield strength, $B=\partial_x A_y-\partial_y A_x$, represents a solenoid-like
magnetic f\/ield at the origin:
\[
B={\pi}\delta(x,y).
\]

\subsection[$SO(2)$ ``Landau model'']{$\boldsymbol{SO(2)}$ ``Landau model''}

Next, we introduce ``Landau model'' on a circle in the presence of magnetic f\/luxes \cite{PRD1983circleQM}
\[
B=I\pi\delta(x,y).
\]
Here, $I$ is the number of magnetic f\/luxes, and takes an integer value.
The corresponding gauge f\/ield is given by
\[
A_x=-\frac{I}{2r^2}y,\qquad A_y=\frac{I}{2r^2}x,
\]
where $r^2=x^2+y^2$.
Since the magnetic f\/luxes are at the origin, the classical motion of a~charged particle on a circle is not af\/fected by the existence of the magnetic f\/luxes.
In quantum mechanics, however, the result is dif\/ferent.
With the covariant derivatives $D_x=\partial_x+iA_x$ and $D_y=\partial_y+iA_y$,
Landau Hamiltonian on 2D plane is given by
\[
H=-\frac{1}{2M}\big(D_x^2+D_y^2\big) =-\frac{1}{2M}\frac{\partial^2}{\partial r^2}+\frac{1}{2Mr^2}\Lambda^2,
\]
where $\Lambda$ is the covariant angular momentum
\[
\Lambda=-ix D_y+iy D_x=-i\frac{\partial}{\partial{\theta}}+A_{\theta},
\]
with $A_{\theta}=\frac{I}{2}$.
On a circle with radius $R$, the Hamiltonian is reduced to $SO(2)$ ``Landau Hamiltonian''
\[
H=\frac{1}{2MR^2}\Lambda^2 =-\frac{1}{2MR^2}\left(\frac{\partial}{\partial\theta}+\frac{i}{2}I\right)^2.
\]
This is a one-dimensional quantum mechanical Hamiltonian easily solved.
In higher dimensions, the covariant angular momentum is not a conserved quantity, but the present case it is, as simply verif\/ied $[\Lambda,H]=0$.
Since the magnetic f\/ield angular momentum does not exist on the circle, the particle angular momentum itself is conserved.
Imposing the periodic boundary condition
\[
u(\theta=2\pi)=u(\theta=0),
\]
the eigenvalue problem is classif\/ied to two cases: even $I$ and odd $I$.
For even $I$, the energy eigenvalue  are given by
\[
E_n=\frac{1}{2MR^2}n^2,
\]
where $n=0,1,2,3, \dots$. The  eigenstates are
\[
u(\theta)_{\pm n}=\frac{1}{\sqrt{2\pi}}e^{i\big(\pm n-\frac{I}{2}\big)\theta}.
\]
Except for the lowest energy level $n=0$, every excited energy level is two-fold degenerate. The physical origin of the double degeneracy comes from  the left and the right movers on the circle. The energy levels are identical to those of free particle on a circle. Thus, for even $I$,
 the magnetic f\/lux  does not af\/fect the energy spectrum of the system.
Meanwhile, for odd $I$, the energy eigenvalues become
\begin{gather}
E'_n=\frac{1}{2MR^2}\left(n+\frac{1}{2}\right)^2,
\label{energyspeccir2}
\end{gather}
%%%%%%%%%%%%%%%
and the corresponding eigenstates are
\[
u'(\theta)_{\pm n}=\frac{1}{\sqrt{2\pi}}e^{i\big(\pm \big(n+\frac{1}{2}\big)-\frac{I}{2}\big)\theta}.
\]
All the Landau levels are doubly degenerate even for $n=0$.
The energy spectra (\ref{energyspeccir2}) are dif\/ferent from those of the free particle on a circle, ref\/lecting the particular role  of  the gauge f\/ield in quantum mechanics.
It is also noted that the number of the magnetic f\/luxes $I$ has nothing to do with the degeneracy in Landau levels unlike the other Landau models (see equation~(\ref{degeneso3n}) for instance).

\section[Generalized 1st Hopf map and $SU(k+1)$ Landau model]{Generalized 1st Hopf map and $\boldsymbol{SU(k+1)}$ Landau model}
\label{sectioncomplexprojective}

\subsection[$SU(k+1)$ Landau model and fuzzy complex projective space]{$\boldsymbol{SU(k+1)}$ Landau model and fuzzy complex projective space}\label{sectioncomplexprojective1}

The projection from sphere to complex projective space,
\[
S^{2k+1} \overset{S^{1}}\longrightarrow \mathbb{C}P^{k},
\]
 is a straightforward generalization of the 1st Hopf map.
(Note $\mathbb{C}P^1 \simeq S^2$.)
In~\cite{Karabali2002}, Karabali and Nair introduced
$SU(k+1)$ Landau model on  $\mathbb{C}P^{k}$
 in $U(1)$ monopole background. Since the $SU(k+1)$ Landau models are reviewed in~\cite{arXiv:hep-th/0407007,hep-th/0606161}, we survey the main results.   The $SU(k+1)$ Landau model Hamiltonian is given by
\begin{gather}
H=\frac{1}{2MR^2}\left(C_{SU(k+1)}-\frac{k}{2(k+1)}I^2\right),
\label{suk+1hamillandau}
\end{gather}
where $C_{SU(k+1)}$ represents the Casimir operator of the $SU(k+1)$ group.
One may notice analogies between (\ref{suk+1hamillandau}) and the spherical Landau model Hamiltonians (see for instance~(\ref{so3landauhamidef})).
The $SU(k+1)$ group has $k$ Casimirs, and hence its irreducible representations are specif\/ied with Young tableau index, $[\lambda_1,\lambda_2,\dots,\lambda_k]$.
In the $SU(k+1)$ Landau model,  Young tableau index is chosen as $[\lambda_1,\lambda_2, \dots,\lambda_k]=[p+q,q, \dots,q]$.
When $q=0$, the Young tableau index becomes  $[p,0, \dots,0]$ representing the fully symmetric representation of the $SU(k+1)$ spinor.  Meanwhile, when $p=0$, the index becomes $[q,q, \dots, q]$ representing the fully symmetric representation of the $SU(k+1)$ complex spinor.
The ``dif\/ference'' between the fundamental  and  complex representations is specif\/ied by $I=p-q$, which corresponds to the $U(1)$ monopole charge. The Landau level index can be taken as $q=n$, since, in LLL ($n=0$), the irreducible representation is reduced to the fully symmetric spinor representation. Thus, the $SU(k+1)$ Young tableau index $(n,I)\equiv[2n+I,n,\dots,n]$ specif\/ies the states of the $n$th Landau level in $U(1)$ monopole background with magnetic charge $\sqrt{\frac{k}{2(k+1)}}I$.
Then, the  energy eigenvalues are derived as
\begin{gather}
E_n=\frac{1}{2MR^2}\left(C_{SU(k+1)}(n,I)-\frac{k}{2(k+1)}I^2\right)
=\frac{1}{2MR^2}\left(n(n+k)+I\left(n+\frac{k}{2}\right)\right) \label{energyeigenvaluecpk}
\end{gather}
and the corresponding degeneracies are
\begin{gather}
d(n,I)=\frac{1}{k!(k-1)!}(2n+I+k)\frac{(I+n+k-1)!(n+k-1)!}{(I+n)!n!}.
 \label{degeneracycpk}
\end{gather}
The $k$ dependence of the LLL energy $E_{\rm LLL}=\frac{I}{4MR^2}k$ is simply understood by remembering that  $\mathbb{C}P^k$ consists of $k$ complex planes each of which equally provides zero-point energy  contribu\-tion~$\frac{B}{2M}=\frac{I}{4MR^2}$ to LLL energy.
The LLL basis elements are constructed by the fully symmetric products of the $SU(k+1)$ Hopf spinor or the Perelomov coherent state of $SU(k+1)$ \cite{Perelomov1972}
\begin{gather}
\psi=
\frac{1}{\sqrt{1+z_1^*z_1+z_2^*z_2+ \cdots+z_{k}^*z_k}}
\begin{pmatrix}
1 \\
z_1\\
z_2\\
\vdots\\
z_k
\end{pmatrix},
\label{sukhopfspinor}
\end{gather}
as
\begin{gather}
\psi_{\rm LLL}^{(m_1,m_2, \dots,m_k)}=
\frac{1}{{\sqrt{1+z_1^*z_1+z_2^*z_2+ \cdots+z_{k}^*z_k}}^I} 1^{m_1}
{z_1}^{m_2}
{z_2}^{m_3}
 \cdots
{z_k}^{m_{k+1}},
\label{LLLbasisSUk}
\end{gather}
where $m_1+m_2+\cdots+m_{k+1}=I$ ($m_1,m_2,\dots,m_k\ge 0$). Thus, the degeneracy in the LLL is  $d_{\rm LLL}=\frac{(I+k)!}{I!k!}$ that coincides with $d(0,I)$ of~(\ref{degeneracycpk}).
With the $SU(k+1)$ Hopf spinor~(\ref{sukhopfspinor}),
the LLL Lagrangian is given by
\[
L_{\rm LLL}=-iI\psi^{\dagger}\frac{d}{dt}\psi.
\]
After quantization, the normalization constraint $\psi^{\dagger}\psi=1$ is imposed on the LLL states:
\[
\psi^t\frac{d}{d\psi}\psi_{\rm LLL}=I\psi_{\rm LLL}.
\]
Indeed, the LLL states (\ref{LLLbasisSUk}) satisfy the above condition.

Fuzzy $\mathbb{C}P^k$ is constructed by taking the fully symmetric representation of  $SU(k+1)$ \cite{arXiv:hep-th/0103023,arXiv:hep-th/0107099,Watamura2005,hep-th/0511114}.
With the $SU(k+1)$ extended Schwinger boson operator
\[
\hat{\psi}=
\begin{pmatrix}
\hat{\psi}_1\\
\hat{\psi}_2\\
\vdots\\
\hat{\psi}_{k+1}
\end{pmatrix}
\]
 satisfying   $[\hat{\psi}_{\alpha},\hat{\psi}_{\beta}]=\delta_{\alpha\beta}$ ($\alpha,\beta=1,2,\dots,k+1$),
coordinates on fuzzy $\mathbb{C}P^k$ are constructed as
\[
\hat{T}_i=\alpha\hat{\psi}^{\dagger}t_i\hat{\psi}
\]
where $t_i$ $(i=1,2, \dots,k^2+2k)$ stand for  the fundamental representation matrix of the $SU(k+1)$ generators, normalized as ${\rm tr}\,(t_it_j)=\frac{1}{2}\delta_{ij}$.
Square of the radius of fuzzy $\mathbb{C}P^k$ is given by
the $SU(k+1)$ Casimir operator:
\begin{gather}
\sum_{i=1}^{k^2+2k}\hat{T}_i \hat{T}_i=\frac{k}{2(k+1)}\alpha^2 I(I+k+1).
\label{formularadiuscpk}
\end{gather}
In particular for fuzzy $\mathbb{C}P^1$ $(k=1)$, equation (\ref{formularadiuscpk})
 reproduces the result of fuzzy two-sphe\-re~(\ref{xsquarequantum}):
\[
\sum_{i=1}^3\hat{T}_i \hat{T}_i=\frac{\alpha^2}{4}I(I+2).
\]
Similarly for fuzzy $\mathbb{C}P^3$ $(k=3)$, equation~(\ref{formularadiuscpk})  becomes
\[
\sum_{i=1}^{15}\hat{T}_i \hat{T}_i=\frac{3}{8}\alpha^2 I(I+4),
\]
which is equal to the Casimir of fuzzy four-sphere
\[
\frac{1}{4}\sum_{a=1}^5 \hat{X}_a\hat{X}_a+\sum_{a<b}\hat{X}_{ab}\hat{X}_{ab}=\frac{3}{8}\alpha^2 I(I+4),
\]
where $\hat{X}_a$ and $\hat{X}_{ab}$ are  given by equations~(\ref{so5gammaop}) and (\ref{so5geneop}), respectively.

The symmetric representation is constructed as
\[
|m_1,m_2, \dots,m_k\rangle_{sym.}=\frac{1}{\sqrt{m_1!m_2!  \cdots m_{k+1}!}}(\hat{\psi}_1^{\dagger})^{m_1}(\hat{\psi}_2^{\dagger})^{m_2} \cdots(\hat{\psi}_{k+1}^{\dagger})^{m_{k+1}}|0\rangle,
\]
where $m_1+m_2+\cdots+m_{k+1}=I$ $(m_1,m_2,\dots,m_{k+1}\ge 0)$.
The number of the states consisting of fuzzy $\mathbb{C}P^k$ is  $d(0,I)=\frac{(I+k)!}{k!I!}$. From the above discussion, the correspondences between the $SU(k+1)$ Landau model and the $\mathbb{C}P^k$ fuzzy manifold is apparent.

\subsection{Relations to spherical Landau models}\label{sectioncomplexprojective2}

As easily verif\/ied, the $SU(2)$ Landau model is equivalent to the $SO(3)$ Landau model.
From the formulae (\ref{energyeigenvaluecpk}) and (\ref{degeneracycpk}), the energy eigenvalues and degeneracies of the $SU(2)$ Landau Hamiltonian are derived as
\[
E_n=\frac{1}{2MR^2}\left(n(n+1)+I\left(n+\frac{1}{2}\right)\right),\qquad
 d_n=2n+I+1.%\label{degeneracyinso3landaumodel2}
\]
These indeed reproduce the results of the $SO(3)$ Landau model, (\ref{energylevelsphere}) and (\ref{degeneso3n}).
Moreover, the $SU(2)$ coherent state
\[
\psi=\frac{1}{\sqrt{1+z^*z}}
\begin{pmatrix}
1\\
z
\end{pmatrix},
%\label{inhomogecp1}
\]
corresponds to the 1st Hopf spinor (\ref{1sthopfspinornorth}) by the relation{\samepage
\[
 z=\frac{\phi_2}{\phi_1}=\frac{x_1+ix_2}{1+x_3},
\]
 and
the $SU(2)$ LLL basis elements (\ref{LLLbasisSUk}) for $k=1$
are equal to equation~(\ref{LLLbasistwosphere}).}

There also exists correspondence between the $SU(4)$  and $SO(5)$ Landau models in the LLL.
Again from equations~(\ref{energyeigenvaluecpk}) and (\ref{degeneracycpk}), the energy eigenvalues and degeneracies of the $SU(4)$ Landau model are read as
%%%%%%%%%%%%%%%%%%%%%%%%%%
\begin{gather}
E_n=\frac{1}{2MR^2}\left(n(n+3)+I\left(n+\frac{3}{2}\right)\right),\nonumber\\
d_n=\frac{1}{12}(n+1)(n+2)(I+n+1)(I+n+2)(I+2n+3).\label{cp3degeneracy}
\end{gather}
The $SU(4)$ Landau level energy is dif\/ferent from that of the $SO(5)$ Landau model~(\ref{SO5energyegenvalues}), only by the total energy shift $\frac{1}{4MR^2}I$. This discrepancy is understood by  a simple geometrical argument~\cite{Karabali2002}: Since $\mathbb{C}P^3\approx S^4\times S^2$,  the $\mathbb{C}P^3$ is regarded as $S^2$-f\/ibration over $S^4$, and the zero-point energy from the extra $S^2$-space, $\frac{B}{2M}=\frac{1}{4MR^2}I$, gives rise to the dif\/ference.
Gene\-rally, the de\-gene\-racy of Landau levels of the $SU(4)$ Landau model~(\ref{cp3degeneracy}) is dif\/ferent from that of  the $SO(5)$ Landau model (\ref{nthdegeneracyonfoursphere}), but
 in the LLL, both quantities coincide to yield $d_{\rm LLL}=\frac{1}{6}(I+1)(I+2)(I+3)$. This manifests the equivalence between the $SO(5)$ and $SU(4)$ Landau models in LLL.
By comparing the 2nd Hopf spinor~(\ref{explicitnorth2ndhopf}) with the
$SU(4)$ coherent state
\[
\psi=
\frac{1}{\sqrt{1+z_1^*z_1+z_2^*z_2+z_3^*z_3}}
\begin{pmatrix}
1 \\
z_1\\
z_2\\
z_3
\end{pmatrix},
\]
we f\/ind the correspondence
\begin{gather*}
z_1=\frac{\psi_2}{\psi_1}=\frac{\phi_2}{\phi_1},\qquad
z_2=\frac{\psi_3}{\psi_1}=\frac{1}{1+x_5}\left(x_4-ix_3-(ix_1+x_2)\frac{\phi_2}{\phi_1}\right),\nonumber\\
z_3=\frac{\psi_4}{\psi_1}=\frac{1}{1+x_5}\left(-ix_1+x_2+(x_4+ix_3)\frac{\phi_2}{\phi_1}\right),
\end{gather*}
and also for the LLL basis elements, (\ref{SO(5)LLLbases}) and (\ref{LLLbasisSUk}) for $k=3$.

\subsection*{Note added}

 After completion of this work, the author learned the works \cite{Sheikh-Jabbari2004,Sheikh-JabbarTorabian2005}. In the appendices of the papers, the fuzzy spheres ($S_F^2$, $S_F^3$, $S_F^4$ and $S^8_F$) are constructed in the context of embedding in Moyal spaces, with emphasis on relations to the Hopf maps.
 Such constructions are completely consistent with the descriptions in the present paper. The author is grateful to Mohammad M. Sheikh-Jabbari for the information.

 \subsection*{Acknowledgements}

 I would like to thank Yusuke Kimura for collaborations.
 Many crucial ingredients in this review are based on the works with him. I am also indebted to Takehiro Azuma for email correspondence about mathematics of fuzzy spheres.
 Since this article is a review-type, many important works not performed by the author are included.
Hereby, I express my gratitude to the researchers whose works are reviewed in the paper.

%\pdfbookmark[1]{References}{ref}
 \LastPageEnding

\end{document}